\documentclass[preprint,12pt]{elsarticle}
\usepackage[utf8]{inputenc}
\usepackage{amsthm}
\usepackage{amsmath}
\usepackage{amsfonts}
\usepackage{amssymb}
\usepackage{graphicx}
\usepackage{epstopdf}
\usepackage{color}
\usepackage{multirow}
\usepackage{mathtools}
\usepackage{xcolor}
\usepackage{a4wide}
\usepackage{bm}
\usepackage{caption}
\usepackage{subcaption}
\usepackage[
  separate-uncertainty = true,
  multi-part-units = repeat
]{siunitx}
\usepackage{rotating}
\usepackage{verbatim}
\usepackage{lineno,hyperref}
\modulolinenumbers[5]
\usepackage{tikz}
\tikzstyle{noeud} = [circle, draw, fill=white, inner sep=2pt]
\journal{Theoretical Computer Science}
\newtheorem{theorem}{Theorem}[section]
\newtheorem{lemma}{Lemma}[section]

\newdefinition{remark}{Remark}[section]
\newtheorem{observation}{Observation}[section]
\newdefinition{definition}{Definition} \newdefinition{example}{Example}
\newdefinition{prob}{Problem definition}
\usepackage[noend,linesnumbered,ruled,lined]{algorithm2e}
\SetAlFnt{\scriptsize}
\usepackage{amsfonts}
\usepackage{makecell}

\usepackage[usenames,dvipsnames]{pstricks}
\usepackage{pstricks-add}
\usepackage{epsfig}
\usepackage{pst-grad} % For gradients
\usepackage{pst-plot} % For axes
\usepackage[space]{grffile} % For spaces in paths
\usepackage{etoolbox} % For spaces in paths
\makeatletter % For spaces in paths
\patchcmd\Gread@eps{\@inputcheck#1 }{\@inputcheck"#1"\relax}{}{}
\makeatother
\usepackage{mathtools}

\usepackage{graphicx}
\usepackage{tabularx}
\usepackage{textcomp}
\usepackage{xcolor}
\usepackage[labelformat=simple]{subcaption}

\usepackage{booktabs}

\usepackage{pgf,tikz,pgfplots}
\pgfplotsset{compat=1.15}
\usepackage{mathrsfs}
\usepackage{hyperref}
\usepackage{multicol}

  \newtheorem{corollary}{Corollary}

\SetCommentSty{mycommfont}

\journal{arXiv}

\usepackage{a4wide}
\begin{document}
\usetikzlibrary{arrows}
\begin{frontmatter}

\title{Exploration on Highly Dynamic Graphs\footnote{A preliminary version of this work appeared in DISC 2025~\cite{Ashish_2025}.}}

\author[1]{Ashish Saxena}
\author[1]{Kaushik Mondal\footnote{Corresponding Author}}
\affiliation[1]{organization={Department of Mathematics},%Department and Organization
            addressline={Indian Institute of Technology Ropar}, 
            city={Rupnagar},
            postcode={140001}, 
            state={Punjab},
            country={India}}
\begin{abstract}
We study the exploration problem by mobile agents in two prominent models of dynamic graphs: $1$-Interval Connectivity and Connectivity Time. The $1$-Interval Connectivity model was introduced by Kuhn et al.~[STOC 2010], and the Connectivity Time model was proposed by Michail et al.~[JPDC 2014]. Recently, Saxena et al.~[TCS 2025] investigated the exploration problem under both models. In this work, we first strengthen the existing impossibility results for the $1$-Interval Connectivity model. We then show that, in Connectivity Time dynamic graphs, exploration is impossible with $\frac{(n-1)(n-2)}{2}$ mobile agents, even when the agents have full knowledge of all system parameters, global communication, full visibility, and infinite memory. This significantly improves the previously known bound of $n$. Moreover, we prove that to solve exploration with $\frac{(n-1)(n-2)}{2}+1$ agents, $1$-hop visibility is necessary. Finally, we present an exploration algorithm that uses $\frac{(n-1)(n-2)}{2}+1$ agents, assuming global communication, $1$-hop visibility, and $O(\log n)$ memory per agent.
\end{abstract}


\end{frontmatter}

\section{Introduction}
The exploration of graphs by mobile agents is a well-studied problem in distributed computing and has foundational importance in theoretical computer science. Originating from early work by Shannon \cite{shannon1993}, the objective is for mobile agents to collectively visit every node in a given network. Depending on the requirements, the task may involve visiting each node at least once (exploration with termination) or repeatedly over time (perpetual exploration). This problem is not only of theoretical interest but also has practical implications for systems involving autonomous agents, such as robots, software agents, or vehicles, where exploration helps in fault detection, information dissemination, or data collection across the network.

The graph exploration problem has been studied under a wide range of assumptions. These include whether the nodes are uniquely labelled or anonymous, whether agents have distinct identities or are indistinguishable, and the mode of communication or interaction among agents, such as using whiteboards, tokens, face-to-face meetings, or vision-based mechanisms. Variations also arise based on the degree of synchrony among agents (asynchronous, semi-synchronous, or fully-synchronous), the extent of their knowledge about the network, and the amount of memory available to them (refer to \cite{Albers_2000, Cohen_2008, Chalopin_2010, Deng_1999, Pelc_1999, Pelc_2005, Pelc_2008, Pelc_2014, Dobrev_2019}, for a comprehensive overview, refer to \cite{Das_2019}). Despite the diversity in models, most of the prior research is on static graphs, meaning the graph structure remains fixed throughout the exploration. While this assumption works well for traditional networks, where changes typically result from failures but it falls short in capturing the behaviour of today's highly dynamic networks.

The dynamic nature of modern networks presents significant challenges in addressing various algorithmic problems in mobile computing and related domains, as the underlying network topology evolves over time. From the perspective of mobile agents, this means that agents must carry out their tasks in an environment that evolves over time steps. A foundational model capturing such dynamic behaviour was introduced by Kuhn et al. \cite{Kuhn_2010}. In their framework, they defined a stability property known as $T$-Interval Connectivity (for $T \geq 1$), which requires that in every sequence of $T$ consecutive rounds, there exists a stable, connected spanning subgraph, although additional edges may appear or disappear in each round.

The formal description of their model is as follows. Let $V$ be a set of static vertices, $S=\{(u,\,v)\,|\, u,v\in V\}$, where $(u,\,v)$ denotes an edge between $u$ $\&\;v$, and $\mathscr{P}(S)$ be the power set of the set $S$. A synchronous dynamic network is modeled as a dynamic graph $\mathcal{G} = (V, \,E)$, where $V$ is a static set of nodes and $E : \mathbb{N } \rightarrow \mathscr{P}(S)$ is a function that maps a round number $r \in \mathbb{N} \cup \{0\}$ to a set $E(r) \in \mathscr{P}(S)$ of undirected edges. For any round $r \geq 0$, we denote the graph by $\mathcal{G}_r=(V, \, E(r))$. Kuhn et al. \cite{Kuhn_2010} provide the following definition of connectivity in the network.

\begin{definition}
    \cite{Kuhn_2010} ($T$-Interval Connectivity) A dynamic graph $\mathcal{G} = (V, \,E)$ is $T$-Interval Connected for $T \geq 1$ if for all $r \in \mathbb{N} \cup \{0\}$, the static graph $G_{r,\,T} := (V,\, \bigcap_{i=r}^{r+T-1} E(i))$ is connected. The graph is said to be $\infty$-Interval Connected if there is a connected static graph $G'=(V, \,E')$ such that for all $r \in \mathbb{N} \cup \{0\}$, $E' \subseteq E(r)$.
\end{definition}
For $T>1$, the graph can not change arbitrarily as it needs to maintain a stable spanning sub-graph. However, for $T = 1$, this means that the graph is connected in every round, but may change arbitrarily between rounds. Later in 2014, Michail et al. \cite{Michail_2014} introduced another natural and practical definition of connectivity of a possibly disconnected dynamic network that they call Connectivity Time. The authors provide the following definition of connectivity.

\begin{definition}
    \cite{Michail_2014} (Connectivity Time)  The Connectivity Time of a dynamic network $\mathcal{G} = (V, \,E)$ is the minimum $T \in \mathbb{N}$ s.t. for all times $r \in \mathbb{N} \cup \{0\}$ the static graph $G_{r,\,T}:= (V,\, \bigcup_{i=r}^{r+T-1} E(i))$ is connected.
\end{definition}

This model generalizes $T$-Interval Connectivity, but unlike $T$-Interval Connectivity, it allows temporary disconnections. Thus, Connectivity Time is strictly weaker than $T$-Interval Connectivity. In the next section, we discuss the model and problem definition.

\section{Model and problem definition}\label{sec:model}
\noindent \textbf{Dynamic graph model:}
A dynamic network is modeled as a \emph{time-varying graph (TVG)}
$\mathcal{G} = (V, E, T, \rho)$, where $V$ is the set of nodes, $E$ is the set of edges, $T$ is the temporal domain, and $\rho : E \times T \rightarrow \{0,1\}$ is the presence function, indicating whether a given edge is available at a given time. The static graph $G = (V, E)$ is called the \emph{underlying graph} (or \emph{footprint}) of $\mathcal{G}$, where $|V| = n$ and $|E| = m$. For a node $v \in V$, let $E(v) \subseteq E$ denote the set of edges incident to $v$ in the footprint, and define the \emph{static degree} of $v$ as $\deg(v) = |E(v)|$.

Assuming time is discrete, the TVG $\mathcal{G}$ can be viewed as a sequence of static graphs $\mathcal{S}_{\mathcal{G}} = \mathcal{G}_0, \mathcal{G}_1, \ldots$, where each snapshot $\mathcal{G}_r = (V, E_r)$ represents the state of $\mathcal{G}$ at round $r$, with $E_r = \{ e \in E \mid \rho(e, r) = 1 \}$. For a node $v \in V$, let $E_r(v) \subseteq E$ denote the set of edges incident to $v$ in $\mathcal{G}_r$, and define the \emph{dynamic degree} of $v$ as $\deg_r(v) = |E_r(v)|$. Let $\mathcal{G}_r^1, \mathcal{G}_r^2, \ldots, \mathcal{G}_r^l$ denote the $l$ connected components of $\mathcal{G}_r$. Note that if $l = 1$, then $\mathcal{G}_r$ is connected; otherwise, it is disconnected. Let $D_r^i$ denote the diameter of $\mathcal{G}_r^i$ for each $i \in [1,l]$, and define $D_r = \max \{ D_r^i \mid i \in [1,l] \}.$ The \emph{dynamic diameter} of $\mathcal{G}$ is then defined as $\hat{D} = \max_{r \geq 0} D_r.$ Each snapshot $\mathcal{G}_r$ is an unweighted, undirected, anonymous graph, that is, nodes have no unique identifiers. Moreover, $\mathcal{G}_r$ is port labelled: each node $v \in \mathcal{G}_r$ assigns distinct port numbers from the range
$[0, \deg_r(v)-1]$ to its incident edges. For an edge $e(u,v) \in E_r$, the port number at $u$ and the port number at $v$ are independent.
There is no relation between the port labels used in $\mathcal{G}_r$ and
$\mathcal{G}_{r'}$ for $r \neq r'$. Nodes have no local storage. A node is referred to as \underline{hole} in round $r$ if no agent is present at that node, and as \underline{multinode} if two or more agents occupy it at round $r$. In this work, we consider two connectivity models: (i) 1-Interval Connectivity, and (ii) Connectivity Time.

\vspace{0.5cm}
\noindent \textbf{Agent model:}
We consider a team of mobile agents that are initially placed arbitrarily on the nodes of $G$. Each agent has a unique identifier from the range $[1, n^c]$, where $c$ is some constant, and knows only its own ID. An agent at node \( v \) is aware of the ports of node \( v \) and knows \( \deg_r(v) \) at round \( r \). Agents are equipped with memory and execute the algorithm under a fully synchronous scheduler, i.e., in each round $t$, every agent executes a Communicate-Compute-Move (CCM) cycle:
\begin{itemize}
\item \textbf{Communicate:} Agents communicate as per the communication model.
\item \textbf{Compute:} Based on its local view and any received information, the agent performs computation, including deciding whether and where to move.
\item \textbf{Move:} The agent moves through a chosen port or stays idle.
\end{itemize}
The time complexity is measured by the number of synchronous rounds. We define a \underline{configuration} as the dynamic graph $\mathcal{G}_r$ together with the positions of all agents. With a slight abuse of notation, we denote the configuration at round $r$ by $\mathcal{G}_r$. A configuration at round $r$ is called \underline{dispersed} if each node contains at most one agent; otherwise, it is called \underline{undispersed}. For any node $v$ and round $r$, let \underline{$\alpha_r(v)$} denote the number of agents at $v$ at the beginning of round $r$ in $\mathcal{G}_r$, and let \underline{$\beta_r(v)$} denote the number of agents at $v$ at the end of round $r$ in $\mathcal{G}_r$.

\vspace{0.5cm}
\noindent \textbf{Communication model:} In this work, we consider an $\ell_c$-hop communication model, where $\ell_c \in [0, \hat{D}]$. Under this model, an agent at a node $v$ in $\mathcal{G}_r^i$ can send messages to all agents within $\ell_c$ hops from $v$ in $\mathcal{G}_r^i$. When $\ell_c = 0$, this reduces to face-to-face (f-2-f) communication~\cite{Augustine_2018}, meaning agents can only communicate if they are co-located at the same node. On the other hand, when $\ell_c = \hat{D}$, $\ell_c$-hop communication becomes global communication~\cite{Ajay_dynamicdisp, Pelc_2006}, allowing two agents in $\mathcal{G}_r^i$ to exchange messages regardless of their positions in $\mathcal{G}_r^i$.

\vspace{0.5cm}
\noindent \textbf{Visibility model:} In this work, we consider an $\ell_v$-hop visibility model, where $\ell_v \in [0, \hat{D}]$. In $\ell_v$-hop visibility \cite{Agarwalla_2018, Avery_2020}, an agent at a particular node (say $w \in \mathcal{G}_r$) can see the set of nodes, say $S$, that are within all the nodes that are within distance $\ell_v$ from $w$ at round $r$. It can also see the sub-graph of $ \mathcal{G}_r$ induced by the nodes in $S$, including the presence/absence of agents in all the nodes of this sub-graph, but can't see the memory of other agents. If $\ell_v=0$, an agent at a particular graph node (say $v \in \mathcal{G}_r$) can see agents present at node $v$ at round $r$, and port numbers associated with node $v$, but can't see the memory of other agents. If $\ell_v=\hat{D}$ (known as full visibility), an agent at a particular graph node (say $v \in \mathcal{G}_r^i$) can see the complete map of $\mathcal{G}_r^i$.

\begin{prob}
    A node $v$ is visited by round $r$ if at least one agent is at node $v$ at round $t$, where $t\in [0, r]$. An algorithm achieves \textit{exploration} if every node is visited at least once.  An algorithm achieves \textit{perpetual exploration} if every node is visited infinitely often.
\end{prob}

In the next section, we discuss the current status of exploration in dynamic graphs.

\section{Related work}
Exploration of dynamic graphs has been widely studied in centralized settings, where agents have full knowledge of the network’s evolution. Notably, optimal exploration schedules have been analyzed under 1-Interval Connectivity \cite{flocchini2012searching} and extended in subsequent works \cite{Erlebach_2015, erlebach2018faster, erlebach2019two}. Specific topologies such as rings and cactuses have also been explored under $T$-Interval and 1-Interval Connectivity, respectively \cite{ilcinkas2018exploration, ilcinkas2014exploration}.

Distributed exploration, with limited agent knowledge, has received less attention. Probabilistic methods like random walks were introduced in early foundation work \cite{avin2008explore}, while deterministic approaches focus on periodic graphs and carrier models \cite{flocchini2012searching, flocchini2013exploration, ilcinkas2011power, ilcinkas2018exploration}. Perpetual exploration and exploration with termination have been studied in 1-Interval Connected rings using 2 or 3 agents under Fsync and Ssync models \cite{bournat2016self, bournat2017computability, di2020distributed}. Other results include exploration with $O(n)$ agents in toroidal networks \cite{gotoh2018group}, and single-agent strategies with partial foresight \cite{gotoh2019exploration}.  A significant advancement in this area is the work by Gotoh et al. \cite{GOTOH2021}, which investigates the fundamental limits of exploration in time-varying graphs under Fsync and Ssync schedulers. In~\cite{GOTOH2021}, the model differs from ours in several important aspects. The authors study exploration in Temporally Connected and $\ell$-Bounded 1-Interval Connected dynamic graphs. A key difference lies in the port-labelling scheme. In their work, the footprint $G$ is port-labelled, and the outgoing ports at a node $u$ are fixed and correspond to the ports of $u$ in the footprint graph $G$. Specifically, if a port $\lambda$ at node $u$ leads to node $v$ in $G$, then whenever the corresponding edge is present in a snapshot $\mathcal{G}_r$, the same port $\lambda$ always leads to $v$. This assumption enforces local stability of ports across time. In contrast, our model imposes no such restriction on port labelling. There is no relation between the port labels of $\mathcal{G}_r$ and those of $\mathcal{G}_{r'}$ for $r \neq r'$. Consequently, a port $\lambda$ at node $u$ that leads to node $v$ in $\mathcal{G}_r$ may lead to a different node $v_1 \neq v$ in a later snapshot $\mathcal{G}_{r'}$. This distinction has direct consequences for agent movement. In~\cite{GOTOH2021}, an agent that decides to move via a port $\lambda$ may fail to traverse the corresponding edge if that edge is absent in the current round; however, this information is always useful to the agents. In temporally connected dynamic graphs, if $\eta$ edges are transient, then $2\eta + 1$ agents can explore $G$, where an edge $e$ is called transient if there exists a round $r$ such that $\rho(e,r_1) = 0$ for every $r_1 \geq r$. According to~\cite{GOTOH2021}, only one agent waits for a missing edge, while the remaining agents use the rotor-router (RR) mechanism. Let $T$ be the time when the $\eta$ transient edges disappear. Under their strategy, at most $2\eta$ agents can get stuck on these edges, while the remaining agent continues executing RR. Since RR is a self-stabilizing mechanism, this agent is still able to explore $G$. In our model, such a situation does not arise. At each round $r$, ports are assigned from the range $[0, \deg_r(v)-1]$, and every port corresponds to an edge that is present in $\mathcal{G}_r$. Thus, although an agent in our model always moves through a port $\lambda$, this movement does not allow the agent to determine its destination. Moreover, if an agent decides to wait for port $\lambda$, then this port may not exist in the next round if the dynamic degree becomes smaller than $\lambda$. Even if port $\lambda$ is present, the neighbor associated with port $\lambda$ may change in the next round.

Recently, Saxena et al.~\cite{Saxena_2025} presented a study on the exploration problem in dynamic graphs. The dynamic graph model considered in their work is identical to the one considered in this paper. The authors have studied the problem under different connectivity models such as 1-Interval Connectivity, $T$-Path Connectivity and Connectivity Time. In their model, the authors have a restricted communication model. In \cite{Saxena_2025}, agents can either use global communication or f-2-f communication. However, they have not provided a study for the case where there is $\ell_c$-hop communication. This makes their communication model very restricted. In this work, our model is similar to \cite{Saxena_2025}. However, we relaxed the communication and visibility model in this work. In \cite{Saxena_2025}, the authors provided the following results.

\begin{theorem}\label{thm:imp_Exp_1-Interval}
    \cite{Saxena_2025} A set of $k \leq n-2$ agents can't solve the exploration problem in the dynamic graphs, which hold the 1-Interval Connectivity. This impossibility holds even if agents have infinite memory, full visibility, global communication, and know the parameters $k$, $n$. This proof is valid for $\hat{D}=O(1)$.
\end{theorem}

\begin{theorem}\label{thm:exp-1-hop}
    \cite{Saxena_2025} (For $n\geq 7$) It is impossible to solve the exploration with $n-1$ mobile agents on a 1-Interval Connected dynamic graph when the agents have 1-hop visibility and unlimited memory, but without global communication, unless they start in a dispersed configuration. This impossibility holds for $\hat{D}=n-1$, even when the agents know both $n$ and $k$.
\end{theorem}

\begin{theorem}\label{thm:exp-global}
    \cite{Saxena_2025} (For $n\geq 7$) It is impossible to solve the exploration problem with $n-1$ mobile agents on a 1-Interval Connected dynamic graph when the agents are equipped with global communication and unlimited memory but lack 1-hop visibility, unless they start in a dispersed configuration. This impossibility holds for $\hat{D}=n-1$, even when the agents know both $n$ and $k$.
\end{theorem}

\begin{theorem}\label{thm:time_lower_exp}
    \cite{Saxena_2025} Any algorithm solving the exploration problem in a 1-Interval Connected Dynamic graph of $n$ nodes requires $\Omega(n)$ rounds even if $\hat{D}=O(1)$. Moreover, this result holds if the agents have infinite memory, are equipped with global communication, have full visibility and know all of $k$, $n$.
\end{theorem}

In Theorem~\ref{thm:exp-1-hop}, the absence of global communication corresponds to face-to-face communication, and in Theorem~\ref{thm:exp-global}, the lack of 1-hop visibility corresponds to 0-hop visibility. It is also important to note that both results hold under the assumption $\hat{D}=n-1$. A natural question is whether Theorems~\ref{thm:exp-1-hop} and~\ref{thm:exp-global} continue to hold when $\hat{D}=O(1)$. We answer this question in the affirmative. In addition, we characterize for which values of $\ell_c$-hop communication, under the assumption that agents have $\ell_v$-hop visibility, exploration is impossible to solve. In \cite{Saxena_2025}, the authors showed that exploration is impossible with at most $n$ agents. However, their result did not yield tight bounds. In this work, we strengthen their impossibility result by proving that exploration is not solvable even with up to $(n-2)(n-1)/2$ agents, and complement this with a matching algorithmic upper bound. Our contributions are presented in the next section.

\section{Our Contributions}

In this work, we present the following results for the exploration problem in $1$-Interval Connectivity and Connectivity Time dynamic graphs.

\begin{enumerate}
    \item We established a strong impossibility result for exploration in $1$-interval connected dynamic graphs (see Theorem~\ref{thm:l-hop-com-l-hop-visibility}). As a consequence, we obtain the following two results (see Remark~\ref{rk:general}).

\begin{itemize}
    \item[(i)] Exploration with $n-1$ mobile agents is impossible when the agents start from an undispersed configuration and are equipped with $1$-hop visibility and $\big(\lceil \hat{D}/2 \rceil - 3\big)$-hop communication. This result does not rely on the assumption $\hat{D} = n-1$, which makes Theorem~\ref{thm:exp-1-hop} more general.
    
    \item[(ii)] Theorem~\ref{thm:exp-global} also holds without the restriction $\hat{D} = n-1$.
\end{itemize}

    \item For the Connectivity Time model, we prove that exploration is impossible with $\frac{(n-2)(n-1)}{2}$ agents, even when the agents are equipped with infinite memory, full visibility, global communication, and complete knowledge of all system parameters (see Theorem~\ref{th:impossibility}).\footnote{Note that if exploration is impossible, then perpetual exploration is also impossible, and if perpetual exploration is possible, then exploration is also possible.}

    \item We further show that exploration remains impossible in Connectivity Time dynamic graphs with $\frac{(n-2)(n-1)}{2}+1$ agents when the agents have infinite memory, $0$-hop visibility, global communication, and complete knowledge of all parameters (see Theorem~\ref{thm:imp_necessary}). This shows that 1-hop visibility is necessary to solve the exploration problem with $\frac{(n-2)(n-1)}{2}+1$ agents.

    \item Finally, we present a perpetual exploration algorithm for Connectivity Time dynamic graphs that uses $\frac{(n-2)(n-1)}{2} + 1$ agents starting from an arbitrary initial configuration. Each agent is equipped with $1$-hop visibility, global communication, and $O(\log n)$ memory (see Theorem~\ref{th:main}).
\end{enumerate}

\section{Impossibility results}
In this section, we provide impossibility results on 1-Interval Connected and Connectivity Time dynamic graphs. 
\subsection{Impossibility results on 1-Interval Connected dynamic graphs}
We now derive a necessary condition to solve exploration in 1-Interval Connected dynamic graphs when agents have $\ell_c$-hop communication and $\ell_v$-hop visibility, with $\ell_c, \ell_v \in [0, \hat{D}]$.
\begin{figure}
    \centering
    \includegraphics[width=0.65\linewidth]{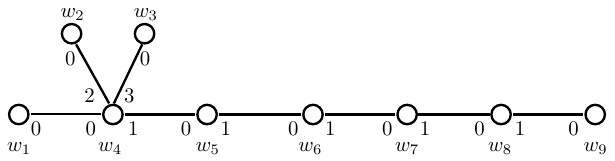}
    \caption{The construction of $G_1$ for $n=9$ and $p=6$.}
    \label{fig:G_1}
\end{figure}
\begin{theorem}\label{thm:l-hop-com-l-hop-visibility}
For $n \geq 7$ and $6 \leq \hat{D} \leq n-1$, the exploration problem on a 1-interval connected graph with $n-1$ mobile agents is impossible to solve under either of the following conditions:

\begin{enumerate}
    \item Agents are equipped with $\ell_c$-hop communication and $\ell_v$-hop visibility such that $\ell_c + \ell_v \leq \lceil \tfrac{\hat{D}}{2} \rceil - 2$, and the initial configuration of agents is undispersed.
    
    \item Agents are equipped with global communication and $0$-hop visibility, and the initial configuration of agents is undispersed.
\end{enumerate}

This impossibility holds even if the agents know $n$ and $k$, the nodes have infinite storage, and the agents have infinite memory.
\end{theorem}

\begin{proof}
    Consider $G$ be a clique of size $n$, and $\hat{D}=p$, where $p\in [6,n-1]$. Since the initial configuration is not dispersed, there exist at least two holes and at least one multinode. Let $V=\{v_1,v_2,\ldots,v_n\}$ be the set of nodes of $G$, where $v_{n-1}$ and $v_n$ are holes, and node $v_1$ is a multinode. Recall $\alpha_r(v)$ denotes the number of agents at node $v$ at the beginning of round $r$. At every round $r$, the adversary constructs $\mathcal{G}_r$ with $D_r=p$. Consider at round $r$ nodes $w_1$, $w_2$, \ldots, $w_n(=v_n)$ be nodes such that $\alpha_r(w_i)\geq \alpha_r(w_{i+1})$ for every $i\in [1,n-1]$ (we show later that in every round $v_n$ remains a hole). Based on $w_i$s, we provide the construction of two graphs $G_1$ and $G_2$.

\vspace{0.2cm}
    \noindent \textbf{Construction of port-labelled graph $G_1$:} Let $P = w_1 \sim w_{n-p+1} \sim w_{n-p+2} \sim \ldots \sim w_{n-1} \sim w_n$. Define $ Q = \{w_2, w_3, \ldots, w_{n-p}\},$ the set of remaining nodes after those used in $P$. For each $w_i \in Q$, add an edge $(w_{n-p+1}, w_i)$. This defines the configuration of $G_1$. An example of $G_1$ for $n=9$ and $p=6$ is shown in Figure \ref{fig:G_1}. The port numbers of $G_1$ are defined as follows. 

    \begin{itemize}
    \item $w_1$ connects to $w_{n-p+1}$ via port $0$.
    \item $w_{n-p+1}$ connects to $w_1$ via port $0$, and to $w_{n-p+2}$ via port $1$.
    \item If $w_i\in Q$, then $w_{n-p+1}$ connects to $w_i$ via port $i+1$, and $w_i$ connects to $w_{n-p+1}$ via port 0.
    \item For $i\in[n-p+2,n-1]$, node $w_i$ connects to $w_{i-1}$ via port $0$, and to $w_{i+1}$ via port $1$.
    \item $w_n$ connects to $w_{n-1}$ via port $0$.
\end{itemize}

\begin{figure}
    \centering
    \includegraphics[width=0.65\linewidth]{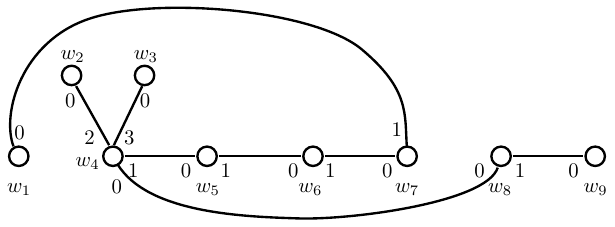}
    \caption{The construction of $G_2$ for $n=9$ and $p=6$.}
    \label{fig:G_2}
\end{figure}
 \noindent \textbf{Construction of port-labelled graph $G_2$:} Let $P = w_1 \sim w_{n-2} \sim \ldots \sim w_{n-p+2}\sim w_{n-p+1} \sim w_{n-1}\sim w_n$. Define $ Q = \{w_2, w_3, \ldots, w_{n-p}\},$ the set of remaining nodes after those used in $P$. For each $w_i \in Q$, add an edge $(w_{n-p+1}, w_i)$. This defines the configuration of $G_2$. An example of $G_2$ for $n=9$ and $p=6$ is shown in Figure \ref{fig:G_2}. The port numbers of $G_2$ are as follows. 

  \begin{itemize}
    \item $w_1$ connects to $w_{n-2}$ via port $0$.
    \item $w_{n-2}$ connects to $w_1$ via port $1$, and to $w_{n-3}$ via port $0$.
    \item For $i\in[n-p+2,n-3]$, node $w_i$ connects to $w_{i-1}$ via port $0$, and to $w_{i+1}$ via port $1$.
    \item $w_{n-p+1}$ connects to $w_{n-1}$ via port 0, and to $w_{n-p+2}$ via port 1. 
    \item If $w_i\in Q$, then $w_{n-p+1}$ connects to $w_i$ via port $i+1$, and $w_i$ connects to $w_{n-p+1}$ via port 0.
    \item $w_{n-1}$ connects to $w_{n-p+1}$ via port 0, and to $w_n$ via port 1.
    \item $w_n$ connects to $w_{n-1}$ via port $0$.
\end{itemize}

It is important to note that the subgraph induced by the node set $S^* = \bigcup_{i=1}^{p-2} \{w_{n-p+i}\}$ in $G_1$ (resp. $G_2$), the positions of agents on the nodes $w_{n-p+i}$ for $i \in [2, p-2]$, and the outgoing port information of each $w_{n-p+i}$ for $i \in [2, p-2]$ are identical in $G_1$ and $G_2$. At the beginning of round $r$, the adversary forms either $G_1$ or $G_2$ (i.e., $\mathcal{G}_r=G_1$ or $G_2$). We claim that if there is no round $r\geq 0$ at which the configuration is dispersed, then the exploration problem cannot be solved by the end of round $r$. At $r=0$, node $w_n(=v_n)$ lies at least two hops away from any agent in $P$, and thus cannot be visited by the end of round $0$. Suppose now that the agents remain in an undispersed configuration up to the start of round $r$. By the construction of $G_1$ (resp. $G_2$), node $w_n(=v_n)$ is at least two hops away from every node occupied by agents. Hence, regardless of how agents move in $G_1$ (resp. $G_2$), node $w_n(=v_n)$ is not visited by the end of round $r$. Therefore, it suffices to show that in every round $r>0$, the adversary can enforce that the configuration at the start of round $r+1$ remains undispersed, thereby preventing the agents from ever being reached to node $w_n$.

    Consider an algorithm $\mathcal{A}$ that solves the exploration problem. If this algorithm fails to achieve a dispersed configuration in some round $r>0$, then by the aforementioned argument, the exploration problem is impossible to solve in round $r+1$. Since the adversary is aware of $G_1$ and the algorithm $\mathcal{A}$, it can pre-compute the outcome of $\mathcal{A}$ at round $r$. If this pre-computation shows that the agents do not achieve the dispersed configuration using $\mathcal{A}$, then at the beginning of round $r$ the adversary constructs $\mathcal{G}_r$ as $G_1$. Otherwise, at the beginning of round $r$, the adversary constructs $\mathcal{G}_r$ as $G_2$. It is important to note that if the dispersed configuration is achieved in $G_1$ at the end of $r$, then $\alpha_r(w_1)=2$ and $\alpha_r(w_i)=1$ for each $i\in [2,n-2]$. If not, then there are at least three holes in $G_1$. Since $\alpha_r(w_i)\geq \alpha_r(w_{i+1})$ for all $i\in[1,n-1]$, it must be the case that $\alpha_r(w_i)=0$ for all $i\in[n-2,n]$. As node $w_{n-2}$ is only reachable via $w_{n-1}$ and $w_{n-3}$, agents can cover at most $w_{n-2}$, leaving at least two holes (namely $w_{n-1}$ and $w_n$) at the end of round $r$. Since there are $n-1$ agents in total, the presence of at least two holes implies that some node must remain a multinode at the end of round $r$. Therefore, if agents achieve dispersed configuration at round $r$, then $\alpha_r(w_1)=2$ and $\alpha_r(w_i)=1$ for each $i\in [2,n-2]$. Now we show that if agents achieve the dispersed configuration in $G_1$, then they do not achieve the dispersed configuration in $G_2$ for the following reasons.

    \begin{enumerate}
        \item When agents have $\ell_c$-hop communication and $\ell_v$-hop visibility satisfying $\ell_c + \ell_v \leq \lceil \tfrac{\hat{D}}{2} \rceil - 2$, and the initial configuration is undispersed, the exploration problem is impossible to solve for the following reason. The agents are equipped with $\ell_c$-hop communication and $\ell_v$-hop visibility, where $\ell_c + \ell_v \leq \lceil \tfrac{\hat{D}}{2} \rceil - 2$. To prove our objective, it suffices to consider the case $\ell_c + \ell_v = \lceil \tfrac{\hat{D}}{2} \rceil - 2$. In terms of visibility, an agent located at node $w_{n-p+\lfloor \frac{p}{2} \rfloor}$ can see the nodes in the set $S$, the positions of the agents on the nodes in $S$, and the outgoing port information of the nodes in $S$, where

    $$S=\bigcup_{i=0}^{\ell_v} \big\{w_{n-p+\lfloor\frac{p}{2}\rfloor-i}, w_{n-p+\lfloor\frac{p}{2}\rfloor+i}\big\}.$$
    
    Since $\ell_v \leq \lceil \tfrac{\hat{D}}{2} \rceil - 2$, we have $S\subseteq S^*$. Therefore, the graph induced by $S$ in $G_1$ (resp. $G_2$), the positions of the agents on the nodes in $S$, and the outgoing port information of the nodes in $S$ are the same in $G_1$ and $G_2$. Therefore, the computation based on $\ell_v$-hop visibility of the agent at node $w_{n-p+\lfloor p/2 \rfloor}$ is the same in both $G_1$ and $G_2$. Therefore, $\ell_v$-hop visibility does not help to break the symmetry for the agent at node $w_{n-p+\lfloor\frac{p}{2}\rfloor}$. The only possibility to break the symmetry for the agent at node $w_{n-p+\lfloor\frac{p}{2}\rfloor}$ is $\ell_c$-hop communication. An agent at node $w_{n-p+\lfloor\frac{p}{2}\rfloor}$ can get the information of agents from set $S_1$, where 

    $$S_1=\bigcup_{i=0}^{\ell_c} \big\{w_{n-p+\lfloor\frac{p}{2}\rfloor-i}, w_{n-p+\lfloor\frac{p}{2}\rfloor+i}\big\}.$$

     Agent at node $w\in S_1$ can share its $\ell_v$-hop visibility with agent at node $w_{n-p+\lfloor\frac{p}{2}\rfloor}$. The $\ell_v$-hop visibility of nodes from set $S_1$ is $S_2$, where 

     $$S_2=\bigcup_{i=0}^{\ell_c+\ell_v} \big\{w_{n-p+\lfloor\frac{p}{2}\rfloor-i}, w_{n-p+\lfloor\frac{p}{2}\rfloor+i}\big\}.$$
    
    Since $\ell_c + \ell_v = \lceil \tfrac{\hat{D}}{2} \rceil - 2$, we have $S_2 \subseteq S^*$. Therefore, the graph induced by $S_2$ in $G_1$ (resp. $G_2$), the positions of the agents on the nodes in $S_2$, and the outgoing port information of the nodes in $S_2$ are the same in $G_1$ and $G_2$. Therefore, $\ell_c$-hop communication does not help to break the symmetry for the agent at node $w_{n-p+\lfloor\frac{p}{2}\rfloor}$ as well. Therefore, the computation of the agent at the node $w_{n-p+\lfloor\frac{p}{2}\rfloor}$ is the same in $G_1$ and $G_2$. Thus, if the agent at node $w_{n-p+\lfloor \frac{p}{2} \rfloor}$ moves to node $w_{n-p+\lfloor \frac{p}{2} \rfloor + 1}$ and achieves a dispersed configuration in $G_1$, then agent at node $w_{n-p+\lfloor \frac{p}{2} \rfloor}$ must also move to node $w_{n-p+\lfloor \frac{p}{2} \rfloor + 1}$ in $G_2$. However, this move results in an undispersed configuration in $G_2$, because to achieve dispersion in $G_2$, the agent at $w_{n-p+\lfloor \frac{p}{2} \rfloor}$ should instead move to $w_{n-p+\lfloor \frac{p}{2} \rfloor - 1}$. Therefore, at the end of round $r$, the dispersed configuration is not achieved. Since in every round $D_r=p$, the value of $\hat{D}$ is $p$.
    \item When agents have global communication and 0-hop visibility, and the initial configuration is undispersed, the exploration problem is impossible to solve for the following reason. Since agents are equipped with 0-hop visibility and global communication, the knowledge of agents in $G_1$ and $G_2$ remains the same. Therefore, the computation of agent at node $w_{n-p+\lfloor p/2\rfloor}$ is identical in $G_1$ and $G_2$. Thus, if the agent at node $w_{n-p+\lfloor \frac{p}{2} \rfloor}$ moves to node $w_{n-p+\lfloor \frac{p}{2} \rfloor + 1}$ and achieves a dispersed configuration in $G_1$, then agent at node $w_{n-p+\lfloor \frac{p}{2} \rfloor}$ moves to node $w_{n-p+\lfloor \frac{p}{2} \rfloor + 1}$ in $G_2$. However, this move results in an undispersed configuration in $G_2$, because to achieve dispersion in $G_2$, the agent at $w_{n-p+\lfloor \frac{p}{2} \rfloor}$ should instead move to $w_{n-p+\lfloor \frac{p}{2} \rfloor - 1}$. Therefore, at the end of round $r$, the dispersed configuration is not achieved. Since in every round $D_r=p$, the value of $\hat{D}$ is $p$.
    \end{enumerate}

     This completes the proof. 
\end{proof}

 Based on Theorem \ref{thm:l-hop-com-l-hop-visibility}, we have the following remark.

\begin{remark}\label{rk:general}
Exploration with $n-1$ mobile agents is impossible when the agents start from an undispersed configuration and are equipped with $1$-hop visibility and $\lceil \hat{D}/2 \rceil - 3$-hop communication. This result does not rely on the assumption $\hat{D} = n-1$. Moreover, Theorem~\ref{thm:l-hop-com-l-hop-visibility} holds for every $\hat{D} \in [6, n-1]$, and thus generalizes the proof of Theorem~\ref{thm:exp-1-hop}.  

Furthermore, Theorem~\ref{thm:exp-global}, which was previously established only for the case $\hat{D} = n-1$, also holds without this restriction.
\end{remark}

\begin{figure}[h]
\centering
 \includegraphics[width=0.55\linewidth]{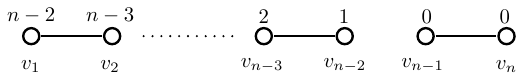}
 \caption{Initial configuration $\mathcal{C}_0$ of $\frac{(n-2)(n-1)}{2}$ agents.}
    \label{fig:imp1}
\end{figure}
\subsection{Impossibility results on Connectivity Time dynamic graphs}\label{sec:imp_Connectivity_Time}
In this section, we establish two results. First, we show that $\frac{(n-2)(n-1)}{2}$ agents are not sufficient to solve the exploration problem. Second, we prove that $\frac{(n-2)(n-1)}{2} + 1$ agents require $1$-hop visibility in order to solve exploration.

\subsubsection{Impossibility results based on the number of agents}\label{sec:imp_Connec_agents}
In this section, we show that exploration is impossible to solve using $\frac{(n-2)(n-1)}{2}$ agents. Let $G$ be a clique of size $n$. Before proceeding with the construction of $\mathcal{G}$, we first outline the high-level idea behind the impossibility result.

\vspace{0.1cm}
\noindent \textbf{High-level idea:} While there remains a non-empty set of unexplored nodes, the goal is to transfer agents from explored and occupied nodes to those unexplored ones. As it can be difficult to differentiate between an unexplored node with an explored but unoccupied node, the algorithm may require to transfer agents from explored and occupied nodes to explored but currently unoccupied nodes as well. If an algorithm succeeds in keeping all explored nodes occupied, then eventually, as the adversary must pick an edge across the cut, some agent will move to an unexplored node and hence the node becomes visited. However, the adversary is powerful: if $b$ agents move from a node with $x$ agents to a node with $y < x$ agents according to some deterministic algorithm, making the new counts $y' = y+b$ and $x' = x-b$, the adversary can flip the roles of these two nodes in the next graph instance, effectively undoing progress as the algorithm performs the reverse operation. The proofs formalize this idea: with fewer than $(n-1)(n-2)/2 + 1$ agents, the adversary can always choose such a flip, whereas with that many agents, it cannot always do that.

\vspace{0.1cm}
\noindent \textbf{Dynamic graph $\mathcal{G}$:} Let $\bm{n=2k}$ for some $k\in \mathbb{N}$, $n\geq 4$ and $T\geq 2$. We give an initial configuration with $\frac{(n-2)(n-1)}{2}$ agents such that the exploration is impossible to solve. Let $v_1$, $v_2$, \ldots, $v_n$ be nodes. At the beginning of round 0, consider $k$ many one length paths as follows: $P_1(=v_1 \sim v_2)$, $P_2(=v_3\sim v_4)$, \ldots, $P_{k-1}(=v_{n-3}\sim v_{n-2})$ and $P_{k}(= v_{n-1}\sim v_n)$. Recall that $\alpha_r(w)$ denotes the number of agents at node $w$ at the beginning of round $r$, and $\beta_r(w)$ denotes the number of agents at node $w$ at the end of round $r$. Let $\alpha_0(v_i)=n-i-1$ for $i\in [1, n-2]$, and $\alpha_0(v_{n-1})=\alpha_0(v_n)=0$ (i.e., nodes $v_{n-1}$ and $v_{n}$ are holes). Let's denote this configuration by $\mathcal{C}_0$ (refer to Figure \ref{fig:imp1}). The total number of agents is $\sum_{i=1}^{n} \alpha_0(v_i)=\sum_{i=1}^{n-2} i=\frac{(n-2)(n-1)}{2}$. At round $r\geq 0$, the adversary maintains $\mathcal{G}_r$ as follows:

\begin{figure}[h]
\centering
\includegraphics[width=1\linewidth]{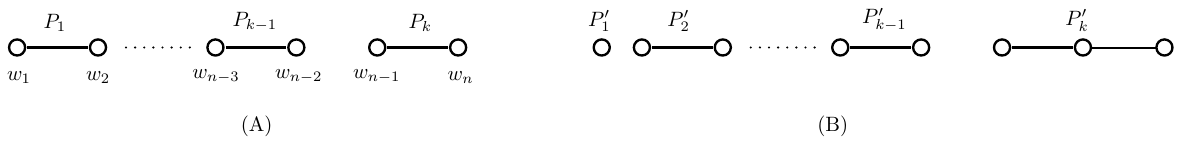}
\caption{(A) Graph $\mathcal{G}_{iT-2}$, (B) Graph $\mathcal{G}_{iT-1}$.}
    \label{fig:imp2}
\end{figure}

\begin{figure}[h]
\centering
\includegraphics[width=1\linewidth]{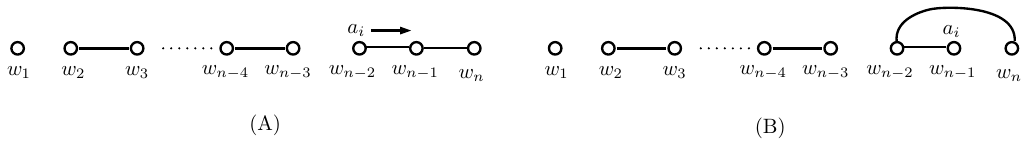}
 \caption{(A) An agent moves from node $w_{n-2}$ to node $w_{n-1}$ at round $r-1$ in $\mathcal{G}_{r-1}$, (B) $\mathcal{G}_{r}$ with respect to $\mathcal{G}_{r-1}$.}
    \label{fig:imp3}
\end{figure}

\begin{figure}[h]
\centering
\includegraphics[width=1\linewidth]{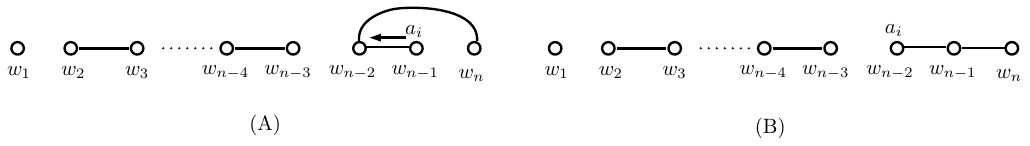}
\caption{(A) An agent moves from node $w_{n-1}$ to node $w_{n-2}$ at round $r-1$ in $\mathcal{G}_{r-1}$, (B) $\mathcal{G}_{r}$ with respect to $\mathcal{G}_{r-1}$.}
    \label{fig:imp4}
\end{figure}

\begin{enumerate}
    \item  $\bm{r\in [0, \,T-2]:}$ It maintains $\mathcal{C}_0$ in these rounds.

         \item $\bm{r \in [iT-1, \, (i+1) T-2]}$, \textbf{where} $\bm{i \geq 1\, \& \,i(}$\textbf{mod} $\bm{2) \neq 0}$: At round $iT-2$, there are $k$ many one length paths, say $P_1(=w_1 \sim w_2)$, $P_2(=w_3\sim w_4)$, \ldots, $P_{k-1}(=w_{n-3}\sim w_{n-2})$, $P_{k}(=w_{n-1}\sim w_n)$ (we separately show that nodes $w_{n-1}$ and $w_n$ are holes at round $iT-2$, with $w_n=v_n$). Note that at the end of round $T-2$, $w_i=v_i$, for every $i$. We can see $\mathcal{G}_{iT-2}$ in Figure \ref{fig:imp2}(A). Based on the movement of agents at round $iT-2$, the adversary forms $k$ paths, say $P_1'$, $P_2'$, \ldots, $P_k'$, at the beginning of round $iT-1$ as follows. 

         If $\beta_{iT-2}(w_1)\geq \beta_{iT-2}(w_2)$, then $P_1'=w_1$. Else, $P_1'=w_2$. For $j\in [2, k-1]$, $P_j'$ is defined as follows.     
         \begin{equation*}
    P_j' =
    \begin{cases} 
        w_{2j-3}\sim w_{2j-1},\;\;  \text{ if } \beta_{iT-2}(w_{2j-3})< \beta_{iT-2}(w_{2j-2}) \text{ and } \beta_{iT-2}(w_{2j-1})\geq \beta_{iT-2}(w_{2j})\\ 
        w_{2j-2}\sim w_{2j-1}, \;\; \text{ if }\beta_{iT-2}(w_{2j-3})\geq \beta_{iT-2}(w_{2j-2}) \text{ and } \beta_{iT-2}(w_{2j-1})\geq \beta_{iT-2}(w_{2j})\\
        w_{2j-3}\sim w_{2j},  \;\; \;\;\;\text{ if } \beta_{iT-2}(w_{2j-3})< \beta_{iT-2}(w_{2j-2}) \text{ and } \beta_{iT-2}(w_{2j-1})< \beta_{iT-2}(w_{2j})\\ 
        w_{2j-2}\sim w_{2j},  \;\;\;\; \;\text{ if } \beta_{iT-2}(w_{2j-3})\geq  \beta_{iT-2}(w_{2j-2}) \text{ and } \beta_{iT-2}(w_{2j-1})< \beta_{iT-2}(w_{2j})\\
    \end{cases}
\end{equation*}
         
If $\beta_{iT-2}(w_{n-3})\geq \beta_{iT-2}(w_{n-2})$, then $P_k'=w_{n-2}\sim w_{n-1}\sim w_n$. Else, $P_k'=w_{n-3}\sim w_{n-1}\sim w_n$. We can see $\mathcal{G}_{iT-1}$ in Figure \ref{fig:imp2}(B).

Without loss of generality, let $P_1'(=w_1)$, $P_2'(=w_2\sim w_3)$, \ldots, $P_{k-1}'(=w_{n-4}\sim w_{n-3})$, $P_{k}'(=w_{n-2}\sim w_{n-1}\sim w_n)$. If $\alpha_{iT-1}(w_{n-2})=0$, the adversary maintains the graph $\mathcal{G}_r$ as $\mathcal{G}_{iT-1}$ for every $r\in [iT, (i+1)T-2]$. If $\alpha_{iT-1}(w_{n-2})>0$, the adversary maintains the graph $\mathcal{G}_r$ for every $r\in [iT, (i+1)T-2]$ as follows.

        If an agent from node $w_{n-2}$ moves to node $w_{n-1}$ at round $r-1$, then adversary at the beginning of round $r$ maintains the following path: $P_1'=w_1$, $P_2'=w_2\sim w_3$,\ldots, $P_{k-1}'=w_{n-4}\sim w_{n-3}$, and $P_{k}'= w_{n-1}\sim w_{n-2}\sim w_{n}$ (refer Figure \ref{fig:imp3}(A) at round $r-1$, and refer Figure \ref{fig:imp3}(B) at round $r$). Otherwise, it maintains the graph $\mathcal{G}_r$ as $\mathcal{G}_{r-1}$. 

        If an agent from node $w_{n-1}$ moves to node $w_{n-2}$ at round $r-1$, then adversary at the beginning of round $r$ maintains the following path: $P_1'=w_1$, $P_2'=w_2\sim w_3$,\ldots, $P_{k-1}'=w_{n-4}\sim w_{n-3}$, and $P_{k}'= w_{n-2}\sim w_{n-1}\sim w_{n}$ (refer Figure \ref{fig:imp4}(A) at round $r-1$, and refer Figure \ref{fig:imp4}(B) at round $r$). Otherwise, it maintains the graph $\mathcal{G}_r$ as $\mathcal{G}_{r-1}$.

        If $i=1$, then it is not difficult to observe that the number of agents at node $\alpha_{T-1}(w_{n-2})\leq 1$ as $\alpha_{T-2}(w_{n-3})+\alpha_{T-2}(w_{n-2})=3$. We show later for every $i$, $\alpha_{iT-1}(w_{n-2})\leq 1$. Thus, the way we change the graph, the agent always stays between node $w_{n-2}$ and $w_{n-1}$ and can not access node $w_n$ in round $r$. We denote this configuration by $\mathcal{C}_{1-2-3}$.

\begin{figure}[h]
\centering
\includegraphics[width=1\linewidth]{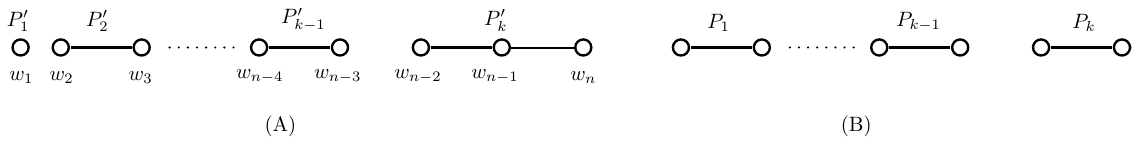}
\caption{(A) Graph $\mathcal{G}_{iT-2}$, (B) Graph $\mathcal{G}_{iT-1}$.}
    \label{fig:imp5}
\end{figure}

      \item $\bm{r \in [iT-1, \, (i+1) T-2]}$, \textbf{where} $\bm{i \geq 1\, \& \,i(}$\textbf{mod} $\bm{2) = 0}$: At the end of round $iT-2$, there are $k$ many paths, say $P_1'(=w_1)$, $P_2'(=w_2\sim w_3)$, \ldots, $P_{k-1}'(=w_{n-4}\sim w_{n-3})$, $P_{k}'=w_{n-2}\sim w_{n-1}\sim w_n$ (we separately show that nodes $w_{n-1}$ and $w_n$ are holes at the beginning of round $iT-2$, with $w_n=v_n$, and at most one agent occupy node $w_{n-2}$). We can see $\mathcal{G}_{iT-2}$ in Figure \ref{fig:imp5}(A). At round $iT-1$, the adversary forms $k$ many one length paths, say $P_1$, $P_2$, \ldots, $P_{k}$. Based on the movement of agents at round $iT-2$, the adversary forms $k$ paths, say $P_1$, $P_2$, \ldots, $P_k$, at the beginning of round $iT-1$ as follows. 

         If $\beta_{iT-2}(w_2)\geq \beta_{iT-2}(w_3)$, then $P_1=w_1\sim w_2$. Else, $P_1=w_1\sim w_3$. For $j\in [2, k-2]$, $P_j$ is defined as follows.
         
         \begin{equation*}
    P_j =
    \begin{cases} 
        w_{2j-2}\sim w_{2j}, \;\;\; \;\; \text{ if } \beta_{iT-2}(w_{2j-2})< \beta_{iT-2}(w_{2j-1}) \text{ and } \beta_{iT-2}(w_{2j})\geq \beta_{iT-2}(w_{2j+1})\\ 
        w_{2j-1}\sim w_{2j}, \;\; \;\;\;\text{ if }\beta_{iT-2}(w_{2j-2})\geq \beta_{iT-2}(w_{2j-1}) \text{ and } \beta_{iT-2}(w_{2j})\geq \beta_{iT-2}(w_{2j+1})\\
        w_{2j-2}\sim w_{2j+1},  \;\; \text{ if } \beta_{iT-2}(w_{2j-2})<\beta_{iT-2}(w_{2j-1}) \text{ and } \beta_{iT-2}(w_{2j})< \beta_{iT-2}(w_{2j+1})\\ 
        w_{2j-1}\sim w_{2j+1},  \;\; \text{ if } \beta_{iT-2}(w_{2j-2})\geq \beta_{iT-2}(w_{2j-1}) \text{ and } \beta_{iT-2}(w_{2j})< \beta_{iT-2}(w_{2j+1})\\
    \end{cases}
\end{equation*}

At the end of round $iT-2$, there is at most one agent in path $P_k'$, and this agent is either at node $w_{n-2}$ or $w_{n-1}$ (we show this separately). If there is no agent in path $P_k'$, then $P_{k-1}$ and $P_k$ are defined as follows. If $\beta_{iT-2}(w_{n-4})\geq \beta_{iT-2}(w_{n-3})$, then $P_{k-1}=w_{n-3}\sim w_{n-2}$ and $P_k=w_{n-1}\sim w_n$. Else, $P_{k-1}=w_{n-4}\sim w_{n-2}$ and $P_k=w_{n-1}\sim w_n$. If there is one agent in path $P_k'$, then $P_{k-1}$ and $P_k$ are defined as follows. Based on agent's position at node $w_{n-2}$ or $w_{n-1}$ at the end of round $iT-2$, the cases are as follows.

\begin{itemize}
    \item If an agent is at node $w_{n-2}$ at the end of round $iT-2$, the path $P_{k-1}$ and $P_k$ are defined as follows. If $\beta_{iT-2}(w_{n-4})\geq \beta_{iT-2}(w_{n-3})$, then $P_{k-1}=w_{n-3}\sim w_{n-2}$ and $P_k=w_{n-1}\sim w_n$. Else, $P_{k-1}=w_{n-4}\sim w_{n-2}$ and $P_k=w_{n-1}\sim w_n$. 
    \item If an agent is at node $w_{n-1}$ at the end of round $iT-2$, the path $P_{k-1}$ and $P_k$ are defined as follows. If $\beta_{iT-2}(w_{n-4})\geq \beta_{iT-2}(w_{n-3})$, then $P_{k-1}=w_{n-3}\sim w_{n-1}$ and $P_k=w_{n-2}\sim w_n$. Else, $P_{k-1}=w_{n-4}\sim w_{n-1}$ and $P_k=w_{n-2}\sim w_n$. 
\end{itemize}

We can see $\mathcal{G}_{iT-1}$ in Figure \ref{fig:imp5}(B). It maintains $\mathcal{G}_r$ as $\mathcal{G}_{iT-1}$ for every $r\in[iT, (i+1)T-2]$. Later, we show that there is no agent in $P_k$, and one of the nodes in $P_k$ is $v_n$. We denote this configuration by $\mathcal{C}_{2-2}$. 
\end{enumerate}

\begin{figure}
    \centering
    \includegraphics[width=0.75\linewidth]{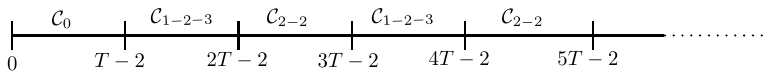}
    \caption{It shows how the agent configurations evolve periodically over the dynamic graph, with specific configurations like $\mathcal{C}_0$, $\mathcal{C}_{1{-}2{-}3}$, and $\mathcal{C}_{2{-}2}$ reappearing at regular intervals.}
    \label{fig:5}
\end{figure}

Figure \ref{fig:5} illustrates the evolution of the configuration over time.
\begin{lemma}\label{lm:corr_connec_time}
    Dynamic graph $\mathcal{G}$ maintains the Connectivity Time property. 
\end{lemma}
\begin{proof}
    For $r\geq 0$, let $\mathcal{G}_r$, $\mathcal{G}_{r+1}$, \ldots, $\mathcal{G}_{r+T-1}$ be consecutive $T$ sequence of graphs, where $\mathcal{G}_i=(V, E(i))$ for $i\in[r,r+T-1]$. Suppose the above dynamic graph $\mathcal{G}$ does not satisfy the Connectivity Time property for some round $r$, i.e., $G_{r, T}:=(V, \cup_{r}^{r+T-1} E(i))$ is not connected. It is important to note that there exists a round $r'$ between $r$ and $r+T-1$ such that $r'=iT-1$, for some $i\in \mathbb{N}$.

    If $i$ is \textbf{odd}, then in each round $t\in [r, iT-2]$, there are $k$ one length paths in $\mathcal{G}_t$: $P_1(=w_1 \sim w_2)$, $P_2(=w_3\sim w_4)$, \ldots, $P_{k-1}(=w_{n-3}\sim w_{n-2})$ and $P_{k}(= w_{n-1}\sim w_{n})$. As per the dynamic graph construction at round $iT-1$, the adversary changes paths as follows: $P_1'(=w_1')$, $P_2'(=w_2'\sim w_3')$, \ldots, $P_{k-1}'(=w_{n-4}'\sim w_{n-3}')$ and $P_{k}'(= w_{n-2}'\sim w_{n-1}'\sim w_n')$, where $w_{2j-1}'\in \{w_{2j-1}, w_{2j}\}$ and $w_{2j}'=\{w_{2j-1}, w_{2j}\}\setminus \{w_{2j-1}'\}$, for every $j\in[1, k]$. Taking the union of the edges from $\mathcal{G}_i$ for $r \leq i \leq r + T - 1$ creates a path of length $n$.

    Similarly, if $i$ is \textbf{even}, then in each round $t\in [r, iT-2]$, there are $k$ paths in $\mathcal{G}_t$: $P_1'(=w_1)$, $P_2'(=w_2\sim w_3)$, ..., $P_{k-1}'(=w_{n-4}\sim w_{n-3})$, and $P_{k}'(= w_{n-2}\sim w_{n-1}\sim w_n)$. By using a similar argument, we can show that the way we modify the construction at round $iT-1$ results in the union of edges from $\mathcal{G}_i$ for $r \leq i \leq r + T - 1$, which forms a path of length $n$. This shows our assumption is wrong. Therefore, this dynamic setting satisfies the Connectivity Time property. This completes the proof.
\end{proof}

We define two notations as follows.
\begin{itemize}
    \item \textbf{(N1)} $\bm{r \in [iT-1, \, (i+1) T-2]}$, \textbf{where} $\bm{i \geq 1\, \& \,i(}$\textbf{mod} $\bm{2) = 0}$ (or $\bm{r\in [0, T-2]}$): In this case, either configuration $\mathcal{C}_0$ or $\mathcal{C}_{2-2}$ is true. Therefore, at round $r$, there are $k$ one length paths $P_1$, $P_2$, \ldots, $P_k$. Let $P_j(=w_{2j-1}\sim w_{2j})$, for every $j \in [1, k]$. If $\alpha_{r}(w_{2j-1})\geq \alpha_r(w_{2j})$, then we denote $w_{2j-1}$ as $w_{2j-1}^r$ and $w_{2j}$ as $w_{2j}^r$. Otherwise, we denote $w_{2j-1}$ as $w_{2j}^r$ and $w_{2j}$ as $w_{2j-1}^r$.

    \vspace{0.2cm}
    \item \textbf{(N2)} $\bm{r \in [iT-1, \, (i+1) T-2]}$, \textbf{where} $\bm{i \geq 1\, \& \,i(}$\textbf{mod} $\bm{2) \neq 0}:$ In this case, configuration $\mathcal{C}_{1-2-3}$ is true. Therefore, at round $r$, there are $k$ paths $P_1'$, $P_2'$, \ldots, $P_k'$. Let $P_j'(=w_{2j-2}\sim w_{2j-1})$, for every $j \in [2, k-1]$, $P_1'(=w_1)$ and $P_k'(=w_{n-2}\sim w_{n-1}\sim w_n)$. If $\alpha_{r}(w_{2j-2})\geq \alpha_r(w_{2j-1})$, then we denote $w_{2j-2}$ as $w_{2j-2}^r$ and $w_{2j-1}$ as $w_{2j-1}^r$. Otherwise, we denote $w_{2j-2}$ as $w_{2j-1}^r$ and $w_{2j-1}$ as $w_{2j-2}^r$. We consider $w_1$ as $w_1^r$, $w_{n-2}$ as $w_{n-2}^r$, $w_{n-1}$ as $w_{n-1}^r$, and $w_n$ as $w_{n}^r$.
\end{itemize} 

\begin{lemma}\label{lm:movement_agents}
The following inequality holds $\forall \;l\in [1, n-2]$ and for every round $r\geq 0$.
\[\sum_{i=1}^{l}\alpha_r(w_i^r) \geq \sum_{i=1}^{l} (n - i - 1)\]
\end{lemma}
\begin{proof}
Proving our lemma is equivalent to showing that the following two inequalities hold for every $j \in [1, k-1]$ at round $r$:

\[
\begin{aligned}
\textbf{(A)} \quad &\sum_{i=1}^{2j} \alpha_r(w_i^r) \geq \sum_{i=1}^{2j} (n - i - 1),\quad \quad\quad\quad\quad
\textbf{(B)} \quad &\sum_{i=1}^{2j-1} \alpha_r(w_i^r) \geq \sum_{i=1}^{2j-1} (n - i - 1).
\end{aligned}
\]

     We use mathematical induction to prove inequalities (A) and (B) for every $r\geq 0$. Both inequalities are true for \( r = 0 \) as $\mathcal{C}_0$ includes \( k \) paths: \( P_1 (= v_1 \sim v_2) \), \( P_2 (= v_3 \sim v_4) \), ..., \( P_{k-1} (= v_{n-3} \sim v_{n-2}) \), and an additional path \( P_k (= v_{n-1} \sim v_n) \). We define \( \alpha_0(v_i) = n - i - 1 \) for \( i \in [1, n-2] \), with \( \alpha_0(v_{n-1}) = \alpha_0(v_n) = 0 \). Therefore, \( w_i^0 = v_i \) for every \( i \leq n-2 \).
     
Assuming the statement is true for round \( r \geq 0 \), we aim to show that it also holds for round \( r + 1 \). There are five possible cases to consider: 
\begin{itemize}
    \item \textbf{Case 1:} $r, r+1\in [0,T-2]$,
    \item \textbf{Case 2:} $r, r+1\in[iT-1, (i+1)T-2]$, where $i$ is an odd number,
    \item  \textbf{Case 3:}  $r, r+1 \in [iT-1, (i+1)T-2]$, where $i$ is an even number,
    \item \textbf{Case 4:}  $r=iT-2$, where $i$ is odd,
    \item \textbf{Case 5:}  $r=iT-2$, where $i$ is even.
\end{itemize}

\vspace{0.2cm}
 \noindent \textbf{Case 1:} In this case, agents are in configuration $\mathcal{C}_0$ at round $r$ and $r+1$.

        \vspace{0.2cm}
         \noindent\textbf{Proof of (A):}  Due to the induction hypothesis, the following inequality holds:
\begin{equation}
\sum_{i=1}^{2j} \alpha_r(w_i^r) \geq \sum_{i=1}^{2j} (n - i - 1)\label{eq:c-1}
\end{equation}
Since the dynamic graph does not change between rounds $0$ and $T-2$, no matter how agents move, for every $j \in [1, k-1]$, we have the following: 
 \begin{equation}
 \alpha_r(w_{2j-1}^{r})+\alpha_r(w_{2j}^{r}) = \alpha_{r+1}(w_{2j-1}^{r+1})+\alpha_{r+1}(w_{2j}^{r+1})\label{eq:c-2}
 \end{equation}
 Therefore, the following inequality holds using Eq. \ref{eq:c-1} and Eq. \ref{eq:c-2}:
\begin{equation}
 \sum_{i=1}^{2j} \alpha_{r+1}(w_i^{r+1})= \sum_{i=1}^{2j} \alpha_r(w_i^r)\geq \sum_{i=1}^{2j} (n - i - 1)\label{eq:c-3}
\end{equation}

\noindent\textbf{Proof of (B):} The inequality $\alpha_r(w_1^r) \geq n - 2$ holds due to the induction hypothesis, and the inequality $\alpha_r(w_1^r) + \alpha_r(w_2^r) \geq (n - 2) + (n - 3)$ is valid due to proof of (A). Therefore, regardless of how the agents move in round $r$, we have $\alpha_{r+1}(w_1^{r+1}) \geq n - 2$ as $\alpha_r(w_1^r) + \alpha_r(w_2^r) = \alpha_{r+1}(w_1^{r+1}) + \alpha_{r+1}(w_2^{r+1})$ and $\alpha_{r+1}(w_1^{r+1}) \geq \alpha_{r+1}(w_2^{r+1})$. This shows (B) holds for $j = 1$. For $j\in[2,k-1]$, we use a contrapositive argument. Suppose for some smallest value $j \geq 2$, the inequality (B) does not hold for round $r+1$. Then the following inequality must be true:
\begin{equation}
\sum_{i=1}^{2j-1} \alpha_{r+1}(w_i^{r+1}) < \sum_{i=1}^{2j-1} (n - i - 1)\label{eq:c-4}
\end{equation}

Due to Eq.~\ref{eq:c-3}, we get:
\begin{equation}
\sum_{i=1}^{2j-2} \alpha_{r+1}(w_i^{r+1}) \geq \sum_{i=1}^{2j-2} (n - i - 1)\label{eq:c-5}
\end{equation}
Therefore, using Eq.~\ref{eq:c-4} and Eq.~\ref{eq:c-5}, the inequality $\alpha_{r+1}(w_{2j-1}^{r+1}) < n - (2j - 1) - 1=n-2j$ holds. Rewrite Eq.~\ref{eq:c-3} as:
\begin{equation}
\alpha_{r+1}(w_{2j}^{r+1}) \geq \sum_{i=1}^{2j} (n - i - 1) - \sum_{i=1}^{2j-1} \alpha_{r+1}(w_i^{r+1})\label{eq:c-6}
\end{equation}
From Eq.~\ref{eq:c-4} and Eq.~\ref{eq:c-6}, the inequality $\alpha_{r+1}(w_{2j}^{r+1}) > n - 2j-1\implies \alpha_{r+1}(w_{2j}^{r+1})\geq n-2j$ holds. Thus, due to our assumption (i.e., Eq. 4), we have $\alpha_{r+1}(w_{2j-1}^{r+1}) < n-2j$ and $\alpha_{r+1}(w_{2j}^{r+1})\geq n-2j$. Therefore, we have $\alpha_{r+1}(w_{2j-1}^{r+1}) < \alpha_{r+1}(w_{2j}^{r+1})$. This leads to a contradiction because, due to (N1), the inequality $\alpha_{r+1}(w_{2j-1}^{r+1}) \geq \alpha_{r+1}(w_{2j}^{r+1})$ holds. This shows that our initial assumption is incorrect. Therefore, inequality (B) holds for round $r+1$.

\vspace{0.2cm}
\noindent \textbf{Case 2:} In this case, round $r, r+1\in[iT-1, (i+1)T-2]$, where $i$ is an odd number, and agents are in configuration $\mathcal{C}_{1-2-3}$. The proof of Eq. (A) and Eq. (B) for round $r+1$ is as follows.

\vspace{0.2cm}
\noindent\textbf{Proof of (B):}  Due to the induction hypothesis, the following inequality holds for $j\geq 1$:
    
\begin{equation}
\sum_{i=1}^{2j-1} \alpha_r(w_i^r) \geq \sum_{i=1}^{2j-1} (n - i - 1)\label{eq:c-7}
\end{equation}

    Since the dynamic graph does not change for every round $r\in[iT-1, (i+1)T-2]$, no matter how agents move, the following holds:
    
    \begin{align}
    \alpha_r(w_1^r)&=\alpha_{r+1}(w_1^r)\label{eq:c-8}\\   \alpha_r(w_{2j}^{r})&+\alpha_r(w_{2j+1}^{r}) = \alpha_{r+1}(w_{2j}^{r+1})+\alpha_{r+1}(w_{2j+1}^{r+1})\;\; \forall j \in [1, k-2]\label{eq:c-9}
    \end{align}

Therefore, inequality (B) holds for round r + 1 using Eq. \ref{eq:c-7}, Eq. \ref{eq:c-8} and Eq. \ref{eq:c-9}.

    \vspace{0.2cm}
\noindent\textbf{Proof of (A):}  We use a contrapositive argument. Suppose for some smallest value $j \geq 1$, the inequality does not hold. Then the following inequality must be true:
    \begin{equation}
        \sum_{i=1}^{2j}\alpha_{r+1}(w_i^{r+1})< \sum_{i=1}^{2j}(n-i-1)\label{eq:c-10}
    \end{equation}

Due to proof of (B), we have:
\begin{equation}
\sum_{i=1}^{2j-1} \alpha_{r+1}(w_i^{r+1}) \geq \sum_{i=1}^{2j-1} (n - i - 1)\label{eq:c-11}
\end{equation}

Therefore, using Eq.~\ref{eq:c-10} and Eq.~\ref{eq:c-11}, the inequality $\alpha_{r+1}(w_{2j}^{r+1}) < n - 2j - 1 \implies \alpha_{r+1}(w_{2j}^{r+1}) \leq n - 2j - 2$ holds. Due to proof of (B), we have:
\begin{equation}
\sum_{i=1}^{2j+1} \alpha_{r+1}(w_i^{r+1}) \geq \sum_{i=1}^{2j+1} (n - i - 1)\label{eq:c-12}
\end{equation}

We now rewrite Eq.~\ref{eq:c-12} as:
\begin{equation}
\alpha_{r+1}(w_{2j+1}^{r+1}) \geq \sum_{i=1}^{2j+1} (n - i - 1) - \sum_{i=1}^{2j} \alpha_{r+1}(w_i^{r+1})\label{eq:c-13}
\end{equation}

From Eq.~\ref{eq:c-10} and Eq.~\ref{eq:c-13}, the inequality $\alpha_{r+1}(w_{2j+1}^{r+1}) > n - 2j-2$ holds. Therefore, due to our assumption (i.e., Eq. \ref{eq:c-10}), we have $\alpha_{r+1}(w_{2j}^{r+1}) \leq n - 2j - 2$ and $\alpha_{r+1}(w_{2j+1}^{r+1}) > n - 2j-2$. This leads to a contradiction because, due to (N2), the inequality $\alpha_{r+1}(w_{2j}^{r+1}) \geq \alpha_{r+1}(w_{2j+1}^{r+1})$ holds. This shows that our initial assumption is incorrect. Therefore, inequality (A) holds for round $r + 1$.

\vspace{0.2cm}
\noindent \textbf{Case 3:} In this case, round $r, r+1\in[iT-1, (i+1)T-2]$, where $i$ is an even number. The proof is similar to Case 1.

\vspace{0.2cm}
 \noindent \textbf{Case 4:} In this case, at round $r = iT - 2$, where $i$ is an odd number, the configuration is either $\mathcal{C}_0$ or $\mathcal{C}_{2\text{-}2}$, and at round $r+1$, it changes to $\mathcal{C}_{1-2-3}$

 \vspace{0.2cm}
\noindent\textbf{Proof of (A):} Due to the induction hypothesis, the following inequality is true:
    \begin{equation}
        \sum_{i=1}^{2j}\alpha_r(w_i^{r})\geq \sum_{i=1}^{2j}(n-i-1)\label{eq:c-14}
    \end{equation}
    The adversary constructs the dynamic graph in round \( r + 1 \) based on the agents' positions at the end of round \( r \). Let $p=\text{max}\{\beta_{r}(w_{2j-1}^r), \beta_{r}(w_{2j}^r)\}$. Therefore, $\alpha_{r+1}(w_{2j-1}^{r+1})=p$ and the value of $\alpha_{r+1}(w_{2j}^{r+1})$ is as follows (recall configuration $\mathcal{C}_{1-2-3}$):
    \begin{equation}
        \alpha_{r+1}(w_{2j}^{r+1})=\text{max}\Big\{\alpha_r(w_{2j-1}^r)+\alpha_r(w_{2j}^r)-p, \text{max}\big\{\beta_r(w_{2j+1}^r),\beta_r(w_{2j+2}^r)\big\}\Big\}\label{eq:c-15}
    \end{equation}
The adversary forms the dynamic graph at round $iT-1$, ensuring the following equality:
    \begin{equation}
        \sum_{i=1}^{2j-1}\alpha_{r+1}(w_i^{r+1})=\sum_{i=1}^{2j-2}\alpha_r(w_i^{r})+p\label{eq:c-16}
    \end{equation}
    Using Eq. \ref{eq:c-16}, the following equality holds.
    \begin{equation}
        \sum_{i=1}^{2j}\alpha_{r+1}(w_i^{r+1})= \sum_{i=1}^{2j-1}\alpha_{r+1}(w_i^{r+1})+ \alpha_{r+1}(w_{2j}^{r+1})\geq\sum_{i=1}^{2j-2}\alpha_r(w_i^{r})+p+\alpha_{r+1}(w_{2j}^{r+1})\label{eq:c-17}
    \end{equation}
    Due to Eq. \ref{eq:c-15}, we have $ \alpha_{r+1}(w_{2j}^{r+1})\geq \alpha_r(w_{2j-1}^r)+\alpha_r(w_{2j}^r)-p$. Thus, using Eq. \ref{eq:c-17}:   
        \begin{equation}
          \sum_{i=1}^{2j}\alpha_{r+1}(w_i^{r+1})\geq \sum_{i=1}^{2j}\alpha_r(w_i^{r})\label{eq:c-18}
        \end{equation}
Using Eq. \ref{eq:c-14} and Eq. \ref{eq:c-18}, the inequality (A) holds for round $r+1$. 

\vspace{0.2cm}
\noindent\textbf{Proof of (B):} For $j=1$, the inequality $\alpha_r(w_1^r)\geq n-2$ holds, and due to the proof of (A), the inequality $\alpha_r(w_1^r)+\alpha_r(w_2^r)\geq n-2+n-3$ holds. Therefore, $\alpha_{r+1}(w_1^{r+1})\geq n-2$ holds regardless of how agents move at round $r$ due to the following reason. If $\alpha_{r+1}(w_1^{r+1})<n-2$, then $\beta_r(w_1^r)<n-2$ and $\beta_r(w_2^r)<n-2$ holds as per the dynamic graph construction at round $iT-1$, $\alpha_{r+1}(w_1^{r+1})=\text{max}\left\{\beta_r(w_1^r),\beta_r(w_2^r)\right\}$. 
Therefore, $\beta_r(w_1^r)\leq n-3$ and $\beta_r(w_2^r)\leq n-3$, and hence the inequality $\beta_r(w_1^r)+\beta_r(w_2^r)\leq 2(n-3)$ holds. Since $\beta_r(w_1^r)+\beta_r(w_2^r)=\alpha_r(w_1^r)+\alpha_r(w_2^r)$, the inequality $\alpha_r(w_1^r)+\alpha_r(w_2^r)\leq 2(n-3)$. This leads to a contradiction as $\alpha_r(w_1^r)+\alpha_r(w_2^r)\geq n-2+n-3$. Therefore, our assumption is wrong, and it implies $\alpha_{r+1}(w_1^{r+1})\geq n-2$. Now we prove (B) when $j\geq 2$. Due to (A), the following inequalities hold for $j\geq 2$.
\begin{equation}
\begin{array}{rl}
\displaystyle \sum_{i=1}^{2j-2}\alpha_r(w_i^{r}) \geq \sum_{i=1}^{2j-2}(n-i-1)\;\;\; \text{and}&
\displaystyle \sum_{i=1}^{2j}\alpha_r(w_i^{r}) \geq \sum_{i=1}^{2j}(n-i-1)\label{eq:c-19}
\end{array}
\end{equation}
In Eq. \ref{eq:c-16}, the lower bound of $p$ is $\left\lceil \frac{\alpha_r(w_{2j-1}^r)+\alpha_r(w_{2j}^r)}{2}\right\rceil$. Using Eq. \ref{eq:c-16}, we have:
\begin{equation}
    \sum_{i=1}^{2j-1}\alpha_{r+1}(w_i^{r+1})\geq\sum_{i=1}^{2j-2}\alpha_r(w_i^{r})+\left\lceil \frac{\alpha_r(w_{2j-1}^r)+\alpha_r(w_{2j}^r)}{2}\right\rceil\label{eq:c-20}
\end{equation}
Using Eq. \ref{eq:c-19}, we have (by taking the sum of inequalities of Eq. \ref{eq:c-19}:
 \begin{equation*}
     2\left( \sum_{i=1}^{2j-2}\alpha_r(w_i^{r})\right)+ \alpha_r(w_{2j-1}^r)+\alpha_r(w_{2j}^r)\geq 2\left(\sum_{i=1}^{2j-2}(n-i-1)\right)+ (n-2j) + (n-2j-1)
 \end{equation*}
 \begin{equation}
    \implies \sum_{i=1}^{2j-2}\alpha_r(w_i^{r})+ \frac{\alpha_r(w_{2j-1}^r)+\alpha_r(w_{2j}^r)}{2}\geq \sum_{i=1}^{2j-2}(n-i-1)+ n-2j-\frac{1}{2}\label{eq:c-21}
 \end{equation}
Since $\lceil x \rceil\geq x$ for every $x\in \mathbb{R}$, the following holds:
\begin{equation}
     \sum_{i=1}^{2j-2}\alpha_r(w_i^{r})+ \left\lceil\frac{\alpha_r(w_{2j-1}^r)+\alpha_r(w_{2j}^r)}{2}\right\rceil\geq \sum_{i=1}^{2j-2}(n-i-1)+ n-2j=\sum_{i=1}^{2j-1} (n-i-1)\label{eq:c-22}
 \end{equation}
Due to Eq. \ref{eq:c-20} and Eq. \ref{eq:c-22}, inequality (B) holds for round $r+1$.

\vspace{0.2cm}
 \noindent \textbf{Case 5:} In this scenario, let \( r = iT - 2 \), where \( i \) is an even integer. At round \( r \), the configuration is \( \mathcal{C}_{1-2-3} \). As per (N2), there are \( k \) paths: \( P_1' (= w_1^r) \), \( P_2' (= w_2^r \sim w_3^r) \), \ldots, \( P_{k-1}' (= w_{n-4}^r \sim w_{n-3}^r) \), and \( P_k' (= w_{n-2}^r \sim w_{n-1}^r \sim w_n^r) \). At round \( r+1 \) as per (N1), there are \( k \) paths: \( P_1 (= w_1^{r+1} \sim w_2^{r+1}) \), \( P_2 (= w_3^{r+1} \sim w_4^{r+1}) \), \ldots, \( P_{k-1} (= w_{n-3}^{r+1} \sim w_{n-2}^{r+1}) \), and \( P_k (= w_{n-1}^{r+1} \sim w_n^{r+1}) \). It holds that \( \alpha_r(w_{2j-1}^{r+1}) \geq \alpha_r(w_{2j}^{r+1}) \) for every \( j \in [1, k-1] \).

    \vspace{0.2cm}
   \noindent\textbf{Proof of (B):}  Due to the induction hypothesis, the following inequality is true:
    \begin{equation}
        \sum_{i=1}^{2j-1}\alpha_r(w_i^{r})\geq \sum_{i=1}^{2j-1}(n-i-1)\label{eq:c-23}
    \end{equation}

    Since $\alpha_{r+1}(w_1^r)=\text{max}\left\{\alpha_r(w_1^r), \text{max}\{\beta_r(w_2^r), \beta_r(w_3^r)\}\right\}$, $\alpha_{r+1}(w_1^r)\geq \alpha_r(w_1^r)\geq n-2$. Therefore, for $j=1$, the inequality (A) holds at round $r+1$. For $j\geq 2$, the inequality (A) holds at round $r+1$ for the following reason. Let $p=\text{max}\{\beta_{r}(w_{2j-2}^r), \beta_{r}(w_{2j-1}^r)\}$ for $j\geq 2$. Therefore, the value of $\alpha_{r+1}(w_{2j-1})$ is as follows.
    \begin{equation}
        \alpha_{r+1}(w_{2j-1}^{r+1})=\text{max}\Big\{\alpha_r(w_{2j-2}^r)+\alpha_r(w_{2j-1}^r)-p, \text{max}\big\{\beta_r(w_{2j}^r),\beta_r(w_{2j+1}^r\big\}\Big\}\label{eq:c-24}
    \end{equation}
The adversary forms the dynamic graph at round $iT-1$, ensuring the following equality is true.
    \begin{equation}
        \sum_{i=1}^{2j-2}\alpha_{r+1}(w_i^{r+1})=\sum_{i=1}^{2j-3}\alpha_r(w_i^{r})+p\label{eq:c-25}
    \end{equation}
    Using Eq. \ref{eq:c-23} and Eq. \ref{eq:c-25}, the following equality holds.
    \begin{equation}
        \sum_{i=1}^{2j-1}\alpha_{r+1}(w_i^{r+1})= \sum_{i=1}^{2j-2}\alpha_{r+1}(w_i^{r+1})+ \alpha_{r+1}(w_{2j-1}^{r+1})\geq \sum_{i=1}^{2j-3}\alpha_r(w_i^{r})+p+\alpha_{r+1}(w_{2j-1}^{r+1})\label{eq:c-26}
    \end{equation}

    Due to Eq. \ref{eq:c-24}, $\alpha_{r+1}(w_{2j-1}^{r+1})\geq\alpha_r(w_{2j-2}^r)+\alpha_r(w_{2j-1}^r)-p$. Therefore, we have the following from Eq. \ref{eq:c-26}.

    \[
        \sum_{i=1}^{2j-1}\alpha_{r+1}(w_i^{r+1})\geq \sum_{i=1}^{2j-3}\alpha_r(w_i^{r})+p+\alpha_{r+1}(w_{2j-1}^{r+1})\geq \sum_{i=1}^{2j-3}\alpha_r(w_i^{r})+p+\alpha_r(w_{2j-2}^r)+\alpha_r(w_{2j-1}^r)-p
    \]
    \begin{equation}
        \implies \sum_{i=1}^{2j-1}\alpha_{r+1}(w_i^{r+1})\geq\sum_{i=1}^{2j-1}\alpha_{r}(w_i^{r})\label{eq:c-27}
    \end{equation}
 Using Eq. \ref{eq:c-23} and Eq. \ref{eq:c-27}, the inequality (B) holds at round $r+1$.

\vspace{0.2cm}
 \noindent\textbf{Proof of (A):}  Due to the proof of (B), the following inequalities are true for any $j\geq 1$.
\begin{equation}
    \sum_{i=1}^{2j-1}\alpha_r(w_i^{r})\geq \sum_{i=1}^{2j-1}(n-i-1)\label{eq:c-28}
\end{equation}
\begin{equation}
    \sum_{i=1}^{2j+1}\alpha_r(w_i^{r})\geq \sum_{i=1}^{2j+1}(n-i-1)\label{eq:c-29}
\end{equation}

Due to Eq. \ref{eq:c-25}, the following inequality is true for $j\geq 1$.
\begin{equation}
    \sum_{i=1}^{2j}\alpha_{r+1}(w_i^{r+1})=\sum_{i=1}^{2j-1}\alpha_r(w_i^{r})+ p\label{eq:c-30}
\end{equation}
The lower bound of $p$ is $\left\lceil \frac{\alpha_r(w_{2j}^r)+\alpha_r(w_{2j+1}^r)}{2}\right\rceil$. Therefore, we get the following inequality from Eq. \ref{eq:c-30}.
\begin{equation}
    \sum_{i=1}^{2j}\alpha_{r+1}(w_i^{r+1})\geq\sum_{i=1}^{2j-1}\alpha_r(w_i^{r})+\left\lceil \frac{\alpha_r(w_{2j}^r)+\alpha_r(w_{2j+1}^r)}{2}\right\rceil\label{eq:c-31}
\end{equation}

 Using Eq. \ref{eq:c-28} and Eq. \ref{eq:c-29}, we get the following inequality (by taking the sum of Eq. \ref{eq:c-28} and Eq. \ref{eq:c-29}:
 \[
     2\left( \sum_{i=1}^{2j-1}\alpha_r(w_i^{r})\right)+ \alpha_r(w_{2j}^r)+\alpha_r(w_{2j+1}^r)\geq 2\left(\sum_{i=1}^{2j-1}(n-i-1)\right)+ (n-2j-2) + (n-2j-1)
 \]
 \[
    \implies \sum_{i=1}^{2j-1}\alpha_r(w_i^{r})+ \frac{\alpha_r(w_{2j}^r)+\alpha_r(w_{2j+1}^r)}{2}\geq \sum_{i=1}^{2j-1}(n-i-1)+ \frac{2n-4j-3}{2}
 \]
 \begin{equation}
    \implies \sum_{i=1}^{2j-1}\alpha_r(w_i^{r})+ \frac{\alpha_r(w_{2j}^r)+\alpha_r(w_{2j+1}^r)}{2}\geq \sum_{i=1}^{2j-1}(n-i-1)+ n-2j-1-\frac{1}{2}\label{eq:c-32}
 \end{equation}
 
We know that the following inequality is true.
 \begin{equation}
     \sum_{i=1}^{2j-1}\alpha_r(w_i^{r})+\left\lceil \frac{\alpha_r(w_{2j}^r)+\alpha_r(w_{2j+1}^r)}{2}\right\rceil\geq    \sum_{i=1}^{2j-1}\alpha_r(w_i^{r})+ \frac{\alpha_r(w_{2j}^r)+\alpha_r(w_{2j+1}^r)}{2}\label{eq:c-33}
 \end{equation}
Due to Eq. \ref{eq:c-32} and Eq. \ref{eq:c-33}, the following holds.
\begin{equation}
     \sum_{i=1}^{2j-1}\alpha_r(w_i^{r})+ \left\lceil\frac{\alpha_r(w_{2j}^r)+\alpha_r(w_{2j+1}^r)}{2}\right\rceil\geq \sum_{i=1}^{2j-1}(n-i-1)+ n-2j-1=\sum_{i=1}^{2j} (n-i-1)\label{eq:c-34}
 \end{equation}
Due to Eq. \ref{eq:c-31} and Eq. \ref{eq:c-34}, the inequality (A) holds at round $r+1$.

    This completes the proof of lemma. 
\end{proof}

\begin{corollary}\label{cor:necessary}
    At round $iT-1$, $\alpha_{iT-1}(w_{n-2}^{iT-1})\leq 1$, where $i$ is an odd number.
\end{corollary}
\begin{proof}
    Due to Lemma \ref{lm:movement_agents}, the following two inequalities hold at round $iT-1$.
\begin{align}
\sum_{j=1}^{n-3} \alpha_{iT-1}(w_j^{iT-1}) \geq \sum_{j=1}^{n-3} (n-j-1) \label{eq:c-35}\\
\sum_{j=1}^{n-2} \alpha_{iT-1}(w_j^{iT-1}) \geq \sum_{j=1}^{n-2} (n-j-1)\label{eq:c-36}
\end{align}
    Therefore, due to Eq. \ref{eq:c-35} and \ref{eq:c-36}, $\alpha_{iT-1}(w_{n-2}^{iT-1})\leq 1$. This completes the proof. 
\end{proof}

    \begin{theorem}\label{th:impossibility}
        If the initial configuration contains at least two holes, then a group of \((n-2)(n-1)/2 \) agents cannot solve the exploration problem in dynamic graphs that maintain the Connectivity Time property. This impossibility holds even if agents have infinite memory, full visibility, global communication, and know the parameters $k$, $n(\geq 4)$, $T(\geq2)$.
    \end{theorem}

    \begin{proof}
    Our dynamic graph construction satisfies the Connectivity Time property as established in Lemma \ref{lm:corr_connec_time}. To prove that exploration is impossible, it suffices to show that node $v_n$ remains a hole at every round $r \geq 0$. If $r\in[0, \, T-2]$, node $v_n$ is not accessible to the agents because node $v_n$ is in path $P_k$, where $v_{n-1}$ is a hole. 

    As per Corollary \ref{cor:necessary}, $\alpha_{T-1}(w_{n-2}^{T-1})\leq 1$. Therefore, at the beginning of round $T-1$, there can be at most one agent at node $w_{n-2}$. If there is no agent, then node $v_n$ is not accessible to the agents because node $v_n$ is in path $P_k'$, where $w_{n-2}$, $w_{n-1}$ are holes. If there is an agent at node \( w_{n-2} \) at the beginning of round \( T-1 \), it has the option to move to node \( w_{n-1} \) during round \( r \in [T-1, 2T-3] \). In this scenario, the adversary constructs the following graph at round \( r+1 \) according to our dynamic graph construction method: the adversary maintains the paths \( P_1', P_2', \ldots, P_{k-1}' \) unchanged and modifies \( P_k' \) from \( w_{n-2} \sim w_{n-1} \sim v_n \) to \( w_{n-1} \sim w_{n-2} \sim v_n \). In this manner, at round \( r+1 \), the node \( v_n \) is two hops away from the agent's position. The adversary follows the same procedure; if the agents move in later rounds from node \( w_{n-1} \) to \( w_{n-2} \), this ensures that during rounds \( r \in [T-1, 2T-2] \), the agents consistently remain at nodes \( w_{n-2} \) and \( w_{n-1} \). Consequently, the agent can not reach node \( v_n \) in any of these rounds.

     At the beginning of round $2T-1$, as per our dynamic graph construction, if there was an agent in path $P_k'$ at round $2T-2$ (there is at most one agent due to Corollary \ref{cor:necessary}), it removes such a node from $P_k'$ at the beginning of round $2T-1$. Therefore, as per dynamic graph construction at the beginning of round $2T-1$, the node $v_n$ is in path $P_k=w_{n-1}\sim w_n(=v_n)$, where node $w_{n-1}$ is a hole. The adversary maintains the same configuration at every round $r\in [2T-1, 3T-2]$. Therefore, the agent can not reach node \( v_n \) in any of these rounds. Using Corollary~\ref{cor:necessary_}, we can always say that in $P_k'$ for round $r\in [iT-1, (i+1)T-2]$, the number of agents is at most one. Since node $w_n(=v_n)$ is unexplored by round $3T-2$, at the beginning of round $3T-1$ in $P_k'$, one agent is at node $w_{n-2}$. At round $r=3T-1$, the node $w_n$ remains unexplored because this case is identical to the case for $r\in [T-1,2T-2]$. Therefore, we can extend the same argument for all $r \geq 3T-1$. Therefore, node $v_n$ remains unexplored for every round $r\geq 0$.

     It is important to note that this proof is independent of the agents' power. Therefore, the proof remains valid even if the agents have full visibility, global communication, and know all parameters. This completes the proof.
    \end{proof}

\subsubsection{Impossibility results for the necessary condition}\label{sec:necessary}
In this section, we show that 1-hop visibility is a necessary condition for solving the exploration problem in Connectivity Time dynamic graphs using $\frac{(n-2)(n-1)}{2}+1$ agents. The proof is similar to Section \ref{sec:imp_Connec_agents}. Let $G$ be a clique of size $n$. The construction of $\mathcal{G}$ is as follows.

\begin{figure}[h]
\centering
 \includegraphics[width=0.55\linewidth]{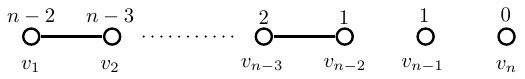}
 \caption{Initial configuration $\mathcal{C}_0'$ of $\frac{(n-2)(n-1)}{2}+1$ agents.}
    \label{fig:nec1}
\end{figure}
\vspace{0.1cm}
\noindent \textbf{Dynamic graph $\mathcal{G}$:} Let $\bm{n=2k}$ for some $k\in \mathbb{N}$, $n\geq 4$ and $T\geq 2$. We give an initial configuration with $\frac{(n-2)(n-1)}{2}+1$ agents such that the exploration is impossible to solve. We assume agents are equipped with global communication and 0-hop visibility, and have knowledge of all parameters. Let $v_1$, $v_2$, \ldots, $v_n$ be nodes. At the beginning of round 0, consider $k+1$ many paths as follows: $P_1(=v_1 \sim v_2)$, $P_2(=v_3\sim v_4)$, \ldots, $P_{k-1}(=v_{n-3}\sim v_{n-2})$ and nodes $P_k(=v_{n-1})$, $P_{k+1}(=v_n)$. Recall that $\alpha_r(w)$ denotes the number of agents at node $w$ at the beginning of round $r$, and $\beta_r(w)$ denotes the number of agents at node $w$ at the end of round $r$. Let $\alpha_0(v_i)=n-i-1$ for $i\in [1, n-2]$, and $\alpha_0(v_{n-1})=1, \alpha_0(v_n)=0$ (i.e., node $v_{n}$ is a hole). Let's denote this configuration by $\mathcal{C}_0'$ (refer to Figure \ref{fig:nec1}). The total number of agents is $\sum_{i=1}^{n} \alpha_0(v_i)=1+\sum_{i=1}^{n-2} i=\frac{(n-2)(n-1)}{2}+1$. At round $r\geq 0$, the adversary maintains $\mathcal{G}_r$ as follows:

\begin{figure}[h]
\centering
\includegraphics[width=1\linewidth]{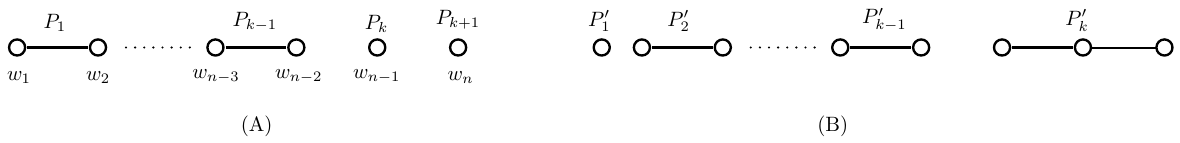}
\caption{(A) Graph $\mathcal{G}_{iT-2}$, (B) Graph $\mathcal{G}_{iT-1}$.}
    \label{fig:nec2}
\end{figure}

\begin{figure}[h]
\centering
\includegraphics[width=0.6\linewidth]{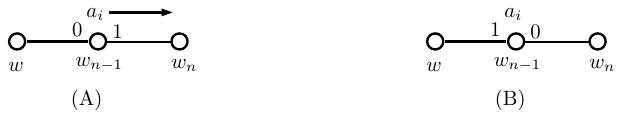}
\caption{In these figures, $w\in \{w_{n-3}, w_{n-2}\}$. (A) As per pre-computation of $P_k'$ an agent moves from node $w_{n-1}$ via port 0, (B) Construction of $P_k'$ at round $iT-1$ if an agent moves from node $w_{n-1}$ via port 0.}
    \label{fig:nec3}
\end{figure}

\begin{figure}[h]
\centering
\includegraphics[width=0.6\linewidth]{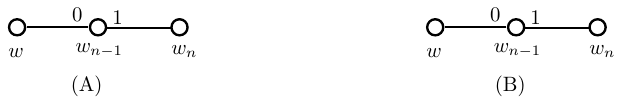}
\caption{In these figures, $w\in \{w_{n-3}, w_{n-2}\}$. (A) As per pre-computation of $P_k'$, no agent moves from node $w_{n-1}$ via port 0, (B) Construction of $P_k'$ at round $iT-1$ if no agent moves from node $w_{n-1}$ via port 0.}
    \label{fig:nec4}
\end{figure}

\begin{enumerate}
    \item $\bm{r\in [0, \,T-2]:}$ It maintains $\mathcal{C}_0'$ in these rounds.

 \item $\bm{r \in [iT-1, \, (i+1) T-2]}$, \textbf{where} $\bm{i \geq 1\, \& \,i(}$\textbf{mod} $\bm{2) \neq 0}$: At round $iT-2$, there are the following paths, say $P_1(=w_1 \sim w_2)$, $P_2(=w_3\sim w_4)$, \ldots, $P_{k-1}(=w_{n-3}\sim w_{n-2})$, $P_{k}(=w_{n-1})$, $P_{k+1}(=w_n)$ (we separately show that nodes $w_{n-1}$ has at most one agent and node $w_n$ is hole at the end of round $iT-2$ for all $i\geq 1$ such that $w_n=v_n$). Note that at the end of round $T-2$, $w_i=v_i$, for every $i$. We can see $\mathcal{G}_{iT-2}$ in Figure \ref{fig:nec2} (A). Based on the movement of agents at round $iT-2$, consider the $k$ paths, say $P_1'$, $P_2'$, \ldots, $P_k'$, at the beginning of round $iT-1$ as follows. 

         If $\beta_{iT-2}(w_1)\geq \beta_{iT-2}(w_2)$, then $P_1'=w_1$. Else, $P_1'=w_2$. For $j\in [2, k-1]$, $P_j'$ is defined as follows.     
         \begin{equation*}
    P_j' =
    \begin{cases} 
        w_{2j-3}\sim w_{2j-1},\;\;  \text{ if } \beta_{iT-2}(w_{2j-3})< \beta_{iT-2}(w_{2j-2}) \text{ and } \beta_{iT-2}(w_{2j-1})\geq \beta_{iT-2}(w_{2j})\\ 
        w_{2j-2}\sim w_{2j-1}, \;\; \text{ if }\beta_{iT-2}(w_{2j-3})\geq \beta_{iT-2}(w_{2j-2}) \text{ and } \beta_{iT-2}(w_{2j-1})\geq \beta_{iT-2}(w_{2j})\\
        w_{2j-3}\sim w_{2j},  \;\; \;\;\;\text{ if } \beta_{iT-2}(w_{2j-3})< \beta_{iT-2}(w_{2j-2}) \text{ and } \beta_{iT-2}(w_{2j-1})< \beta_{iT-2}(w_{2j})\\ 
        w_{2j-2}\sim w_{2j},  \;\;\;\; \;\text{ if } \beta_{iT-2}(w_{2j-3})\geq  \beta_{iT-2}(w_{2j-2}) \text{ and } \beta_{iT-2}(w_{2j-1})< \beta_{iT-2}(w_{2j})\\
    \end{cases}
\end{equation*}
         
If $\beta_{iT-2}(w_{n-3})\geq \beta_{iT-2}(w_{n-2})$, then $P_k'=w_{n-2}\sim w_{n-1}\sim w_n$ such that node $w_{n-1}$ is connected to node $w_{n-2}$ and $w_n$ via port 0 and 1, respectively. Else, $P_k'=w_{n-3}\sim w_{n-1}\sim w_n$ such that node $w_{n-1}$ is connected to node $w_{n-3}$ and $w_n$ via port 0 and 1, respectively. Since the adversary is aware of the agent's algorithm, it can pre-compute the agent's movements in $P_k'$. If there is an agent at node $w_{n-1}$, moves to node $w_n$ via port 1 in $P_k'$ (refer to Figure \ref{fig:nec3} (A)), it changes the port numbers of node $w_{n-1}$. It connects node $w_n$ via port 0, and another node via port 1 (refer to Figure \ref{fig:nec3} (B). Else if no agent from node $w_{n-1}$ moves to node $w_n$ in $P_k'$ (refer to Figure \ref{fig:nec4} (A)), it keeps the same port numbers at node $w_{n-1}$ in $P_k'$ (refer to Figure \ref{fig:nec4} (B).

Based on the pre-computation of agents' movement in $P_k'$ at round $iT-1$, the adversary forms $k$ paths $P_1'$, $P_2'$, \ldots, $P_k'$ at the beginning of round $iT-1$. We can see $\mathcal{G}_{iT-1}$ in Figure \ref{fig:nec2} (B). Without loss of generality, let $P_1'(=w_1)$, $P_2'(=w_2\sim w_3)$, \ldots, $P_{k-1}'(=w_{n-4}\sim w_{n-3})$, $P_{k}'(=w_{n-2}\sim w_{n-1}\sim w_n)$. If $\alpha_{iT-1}(w_{n-2})=\alpha_{iT-1}(w_{n-1})=0$, the adversary maintains the graph $\mathcal{G}_r$ as $\mathcal{G}_{iT-1}$ for every $r\in [iT, (i+1)T-2]$. If $\alpha_{iT-1}(w_{n-2})>0$ or $\alpha_{iT-1}(w_{n-1})>0$, the adversary maintains the graph $\mathcal{G}_r$ for every $r\in [iT, (i+1)T-2]$ as follows (we show later that there are at most two agents in $P_k'$). Since there are at most two agents in $P_k'$ at the end of round $r-1$, it forms $P_k'$ at the beginning of round $r$ as follows:

\begin{itemize}
    \item If $\beta_{r-1}(w_{n-2})>\beta_{r-1}(w_{n-1})$, then it forms $P_k'(=w_{n-2}\sim w_{n-1}\sim w_n)$.
    \item If $\beta_{r-1}(w_{n-2})<\beta_{r-1}(w_{n-1})$, then it forms $P_k'(=w_{n-1}\sim w_{n-2}\sim w_n)$.
    \item If $\beta_{r-1}(w_{n-2})=\beta_{r-1}(w_{n-1})=0$, then it forms $P_k'(=w_{n-2}\sim w_{n-1}\sim w_n)$.
    \item If $\beta_{r-1}(w_{n-2})=\beta_{r-1}(w_{n-1})=1$, then consider $P_k'(=w_{n-2}\sim w_{n-1}\sim w_n)$ such that node $w_{n-1}$ is connected to node $w_{n-2}$ and $w_n$ via port 0 and 1, respectively. Since the adversary is aware of the agent's algorithm, it can pre-compute the agent's movements in $P_k'$. Based on pre-computation, if an agent at node $w_{n-1}$ moves to node $w_n$ via port 1 in $P_k'$, at the beginning of round $r$, it forms $P_k'(=w_{n-2}\sim w_{n-1}\sim w_n)$ such that node $w_{n-1}$ is connected to node $w_{n-2}$ and $w_n$ via port 1 and 0, respectively. Otherwise, at the beginning of round $r$, it forms $P_k'(=w_{n-2}\sim w_{n-1}\sim w_n)$ such that node $w_{n-1}$ is connected to node $w_{n-2}$ and $w_n$ via port 0 and 1, respectively.
\end{itemize}
 We denote this configuration by $\mathcal{C}_{1-2-3}'$.
 
 \vspace{0.2cm}
\begin{figure}[h]
\centering
\includegraphics[width=1\linewidth]{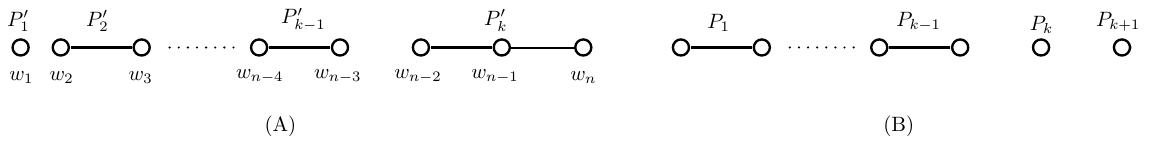}
\caption{(A) Graph $\mathcal{G}_{iT-2}$, (B) Graph $\mathcal{G}_{iT-1}$.}
    \label{fig:nec5}
\end{figure}
 \item $\bm{r \in [iT-1, \, (i+1) T-2]}$, \textbf{where} $\bm{i \geq 1\, \& \,i(}$\textbf{mod} $\bm{2) = 0}$: At the end of round $iT-2$, there are $k$ many paths, say $P_1'(=w_1)$, $P_2'(=w_2\sim w_3)$, \ldots, $P_{k-1}'(=w_{n-4}\sim w_{n-3})$, $P_{k}'=w_{n-2}\sim w_{n-1}\sim w_n$ (we separately show that node $w_n(=v_n)$ is a hole at the end of round $iT-2$, and that there are at most two agents in $P_k'$). We can see $\mathcal{G}_{iT-2}$ in Figure \ref{fig:nec5}(A). At round $iT-1$, the adversary forms $k+1$ many paths, say $P_1$, $P_2$, \ldots, $P_{k+1}$. Based on the movement of agents at round $iT-2$, the adversary forms $k$ paths, say $P_1$, $P_2$, \ldots, $P_k$, at the beginning of round $iT-1$ as follows. 

         If $\beta_{iT-2}(w_2)\geq \beta_{iT-2}(w_3)$, then $P_1=w_1\sim w_2$. Else, $P_1=w_1\sim w_3$. For $j\in [2, k-2]$, $P_j$ is defined as follows.
         
         \begin{equation*}
    P_j =
    \begin{cases} 
        w_{2j-2}\sim w_{2j}, \;\;\; \;\; \text{ if } \beta_{iT-2}(w_{2j-2})< \beta_{iT-2}(w_{2j-1}) \text{ and } \beta_{iT-2}(w_{2j})\geq \beta_{iT-2}(w_{2j+1})\\ 
        w_{2j-1}\sim w_{2j}, \;\; \;\;\;\text{ if }\beta_{iT-2}(w_{2j-2})\geq \beta_{iT-2}(w_{2j-1}) \text{ and } \beta_{iT-2}(w_{2j})\geq \beta_{iT-2}(w_{2j+1})\\
        w_{2j-2}\sim w_{2j+1},  \;\; \text{ if } \beta_{iT-2}(w_{2j-2})<\beta_{iT-2}(w_{2j-1}) \text{ and } \beta_{iT-2}(w_{2j})< \beta_{iT-2}(w_{2j+1})\\ 
        w_{2j-1}\sim w_{2j+1},  \;\; \text{ if } \beta_{iT-2}(w_{2j-2})\geq \beta_{iT-2}(w_{2j-1}) \text{ and } \beta_{iT-2}(w_{2j})< \beta_{iT-2}(w_{2j+1})\\
    \end{cases}
\end{equation*}

At the end of round $iT-2$, there are at most two agents in path $P_k'$, and these agents are at nodes $w_{n-2}$ and $w_{n-1}$ (we show this separately). At the beginning of round $iT-1$, it forms $P_{k-1}$, $P_k$, and $P_{k+1}$ as follows.

\begin{itemize}
    \item If there is no agent in path $P_k'$, then $P_{k-1}$, $P_k$, $P_{k+1}$ are defined as follows. If $\beta_{iT-2}(w_{n-4})\geq \beta_{iT-2}(w_{n-3})$, then $P_{k-1}=w_{n-3}\sim w_{n-2}$, $P_k=w_{n-1}$ and $P_{k+1}=w_n$. Else, $P_{k-1}=w_{n-4}\sim w_{n-2}$, $P_k=w_{n-1}$ and $P_{k+1}=w_n$.
    \item If there is one agent in path $P_k'$, then $P_{k-1}$, $P_k$, $P_{k+1}$ are defined as follows.
    \begin{itemize}
        \item If an agent is at node $w_{n-2}$ at the end of round $iT-2$, the path $P_{k-1}$ and $P_k$ are defined as follows. If $\beta_{iT-2}(w_{n-4})\geq \beta_{iT-2}(w_{n-3})$, then $P_{k-1}=w_{n-3}\sim w_{n-2}$, $P_k=w_{n-1}$ and $P_{k+1}=w_n$. Else, $P_{k-1}=w_{n-4}\sim w_{n-2}$, $P_k=w_{n-1}$ and $P_{k+1}=w_n$. 
        \item If an agent is at node $w_{n-1}$ at the end of round $iT-2$, the path $P_{k-1}$ and $P_k$ are defined as follows. If $\beta_{iT-2}(w_{n-4})\geq \beta_{iT-2}(w_{n-3})$, then $P_{k-1}=w_{n-3}\sim w_{n-1}$, $P_k=w_{n-2}$ and $P_{k+1}=w_n$. Else, $P_{k-1}=w_{n-4}\sim w_{n-1}$, $P_k=w_{n-2}$ and $P_{k+1}=w_n$.
    \end{itemize}
    \item If there are two agents in path $P_k'$, then $P_{k-1}$, $P_k$, $P_{k+1}$ are defined as follows.
    \begin{itemize}
        \item Suppose two agents are at node $w_{n-2}$. If $\beta_{iT-2}(w_{n-4})\geq \beta_{iT-2}(w_{n-3})$, then $P_{k-1}=w_{n-3}\sim w_{n-2}$, $P_k=w_{n-1}$ and $P_{k+1}=w_n$. Else, $P_{k-1}=w_{n-4}\sim w_{n-2}$, $P_k=w_{n-1}$ and $P_{k+1}=w_n$. 
        \item Suppose two agents are at node $w_{n-1}$. If $\beta_{iT-2}(w_{n-4})\geq \beta_{iT-2}(w_{n-3})$, then $P_{k-1}=w_{n-3}\sim w_{n-1}$, $P_k=w_{n-2}$ and $P_{k+1}=w_n$. Else, $P_{k-1}=w_{n-4}\sim w_{n-1}$, $P_k=w_{n-2}$ and $P_{k+1}=w_n$.
        \item Suppose one agents is at node $w_{n-1}$, and other is at node $w_{n-2}$. If $\beta_{iT-2}(w_{n-4})\geq \beta_{iT-2}(w_{n-3})$, then $P_{k-1}=w_{n-3}\sim w_{n-2}$, $P_k=w_{n-1}$ and $P_{k+1}=w_n$. Else, $P_{k-1}=w_{n-4}\sim w_{n-2}$, $P_k=w_{n-1}$ and $P_{k+1}=w_n$.
    \end{itemize}
\end{itemize}

We can see $\mathcal{G}_{iT-1}$ in Figure \ref{fig:nec5} (B). It maintains $\mathcal{G}_r$ as $\mathcal{G}_{iT-1}$ for every $r\in[iT, (i+1)T-2]$. Later, we show that there is no agent in $P_{k+1}$; that is, node $v_n$ is a hole. We denote this configuration by $\mathcal{C}_{2-2}'$. 
\end{enumerate}

\begin{lemma}\label{lm:corr_connec_time_new}
    Dynamic graph $\mathcal{G}$ maintains the Connectivity Time property. 
\end{lemma}
\begin{proof}
    For $r\geq 0$, let $\mathcal{G}_r$, $\mathcal{G}_{r+1}$, \ldots, $\mathcal{G}_{r+T-1}$ be consecutive $T$ sequence of graphs, where $\mathcal{G}_i=(V, E(i))$ for $i\in[r,r+T-1]$. Suppose the above dynamic graph $\mathcal{G}$ does not satisfy the Connectivity Time property for some round $r$, i.e., $G_{r, T}:=(V, \cup_{r}^{r+T-1} E(i))$ is not connected. It is important to note that there exists a round $r'$ between $r$ and $r+T-1$ such that $r'=iT-1$, for some $i\in \mathbb{N}$.

    If $i$ is \textbf{odd}, then in each round $t\in [r, iT-2]$, there are $k+1$ paths in $\mathcal{G}_t$: $P_1(=w_1 \sim w_2)$, $P_2(=w_3\sim w_4)$, \ldots, $P_{k-1}(=w_{n-3}\sim w_{n-2})$, $P_{k}(= w_{n-1})$, $P_{k+1}(= w_{n})$. As per the dynamic graph construction at round $iT-1$, the adversary changes paths as follows: $P_1'(=w_1')$, $P_2'(=w_2'\sim w_3')$, \ldots, $P_{k-1}'(=w_{n-4}'\sim w_{n-3}')$ and $P_{k}'(= w_{n-2}'\sim w_{n-1}'\sim w_n')$, where $w_{2j-1}'\in \{w_{2j-1}, w_{2j}\}$ and $w_{2j}'=\{w_{2j-1}, w_{2j}\}\setminus \{w_{2j-1}'\}$, for every $j\in[1, k]$. Taking the union of the edges from $\mathcal{G}_i$ for $r \leq i \leq r + T - 1$ creates a path of length $n$.

    Similarly, if $i$ is \textbf{even}, then in each round $t\in [r, iT-2]$, there are $k$ paths in $\mathcal{G}_t$: $P_1'(=w_1)$, $P_2'(=w_2\sim w_3)$, ..., $P_{k-1}'(=w_{n-4}\sim w_{n-3})$, and $P_{k}'(= w_{n-2}\sim w_{n-1}\sim w_n)$. By using a similar argument, we can show that the way we modify the construction at round $iT-1$ results in the union of edges from $\mathcal{G}_i$ for $r \leq i \leq r + T - 1$, which forms a path of length $n$. This shows our assumption is wrong. Therefore, this dynamic setting satisfies the Connectivity Time property. This completes the proof.
\end{proof}

We define two notations as follows.
\begin{itemize}
    \item \textbf{(N1)} $\bm{r \in [iT-1, \, (i+1) T-2]}$, \textbf{where} $\bm{i \geq 1\, \& \,i(}$\textbf{mod} $\bm{2) = 0}$ (or $\bm{r\in [0, T-2]}$): In this case, either configuration $\mathcal{C}_0'$ or $\mathcal{C}_{2-2}'$ is true. Therefore, at round $r$, there are $k+1$ one length paths $P_1$, $P_2$, \ldots, $P_{k+1}$. Let $P_j(=w_{2j-1}\sim w_{2j})$, for every $j \in [1, k-1]$, $P_k(=w_{n-1})$ and $P_{k+1}(=w_n)$. We denote $w_{n-1}$ as $w_{n-1}^r$, and $w_n$ as $w_n^r$. For every $j\in [1,k-1]$, if $\alpha_{r}(w_{2j-1})\geq \alpha_r(w_{2j})$, then we denote $w_{2j-1}$ as $w_{2j-1}^r$ and $w_{2j}$ as $w_{2j}^r$. Otherwise, we denote $w_{2j-1}$ as $w_{2j}^r$ and $w_{2j}$ as $w_{2j-1}^r$. 

    \vspace{0.2cm}
    \item \textbf{(N2)} $\bm{r \in [iT-1, \, (i+1) T-2]}$, \textbf{where} $\bm{i \geq 1\, \& \,i(}$\textbf{mod} $\bm{2) \neq 0}:$ In this case, configuration $\mathcal{C}_{1-2-3}$ is true. Therefore, at round $r$, there are $k$ paths $P_1'$, $P_2'$, \ldots, $P_k'$. Let $P_j'(=w_{2j-2}\sim w_{2j-1})$, for every $j \in [2, k-1]$, $P_1'(=w_1)$ and $P_k'(=w_{n-2}\sim w_{n-1}\sim w_n)$. If $\alpha_{r}(w_{2j-2})\geq \alpha_r(w_{2j-1})$, then we denote $w_{2j-2}$ as $w_{2j-2}^r$ and $w_{2j-1}$ as $w_{2j-1}^r$. Otherwise, we denote $w_{2j-2}$ as $w_{2j-1}^r$ and $w_{2j-1}$ as $w_{2j-2}^r$. We consider $w_1$ as $w_1^r$ and $w_n$ as $w_{n}^r$. If $\alpha_r(w_{n-2})\geq \alpha_r(w_{n-1})$, then $w_{n-2}$ as $w_{n-2}^r$, $w_{n-1}$ as $w_{n-1}^r$. Otherwise, $w_{n-1}$ as $w_{n-2}^r$, $w_{n-2}$ as $w_{n-1}^r$.
\end{itemize}

\begin{lemma}\label{lm:condition}
The following inequality holds $\forall \;l\in [1, n-2]$ and for every round $r\geq 0$.
\[\sum_{i=1}^{l}\alpha_r(w_i^r) \geq \sum_{i=1}^{l} (n - i - 1)\]
\end{lemma}
\begin{proof}
The proof is analogous to that of Lemma~\ref{lm:movement_agents}.
\end{proof}

\begin{corollary}\label{cor:necessary_}
    At round $iT-1$, there are at most two agents in $P_k'$, where $i$ is an odd number.
\end{corollary}
\begin{proof}
    Due to Lemma \ref{lm:movement_agents}, the following inequality holds at round $iT-1$.
\begin{equation}
\sum_{j=1}^{n-3} \alpha_{iT-1}(w_i^{iT-1}) \geq \sum_{j=1}^{n-3} (n-j-1) \label{Eq:37}
\end{equation}
    Due to the total number of agents being $\frac{(n-2)(n-1)}{2}+1$ and Eq. \ref{Eq:37}, the remaining agents are at most two, which will be at path $P_k'$. This completes the proof. 
\end{proof}

\begin{theorem}\label{thm:imp_necessary}
        It is impossible to solve the exploration problem with $\frac{(n-2)(n-1)}{2}+1$ mobile agents in the Connectivity Time dynamic graph when the agents are equipped with global communication, 0-hop visibility and unlimited memory. This impossibility holds even when the agents know $n$, $k$ and $T$.
\end{theorem}
\begin{proof}
Our dynamic graph construction satisfies the Connectivity Time property, as shown in Lemma~\ref{lm:corr_connec_time_new}. To prove the impossibility of exploration, it suffices to show that node $v_n$ remains a hole at every round $r \geq 0$.

For rounds $r \in [0, T-2]$, the node $v_n$ is isolated by construction. Hence, it is inaccessible to all agents and therefore remains unexplored. Now consider the dynamic graph construction for rounds $r \in [iT-1, (i+1)T-2]$, where $i$ is odd. Without loss of generality, assume that $\beta_{T-2}(w_{n-3}) \geq \beta_{T-2}(w_{n-2})$. By construction, we define the path $P_k' = w_{n-2} \sim w_{n-1} \sim w_n,$ where $w_{n-2} \in \{v_{n-3}, v_{n-2}\}$, $w_{n-1} = v_{n-1}$, and $w_n = v_n$. By Corollary~\ref{cor:necessary_}, there are two agents in $P_k'$. Since $w_{n-1} = v_{n-1}$ and this node is isolated during the first $T-2$ rounds, it initially contains exactly one agent. Moreover, node $w_n (= v_n)$ contains no agent. Hence, we obtain $\alpha_{T-1}(w_{n-1}) = 1$. Hence, at the beginning of round $T-1$, there are exactly two agents in $P_k'$: one at $w_{n-2}$ and one at $w_{n-1}$. Suppose, the node $w_{n-1}$ is connected to $w_{n-2}$ and $w_n$ via ports $0$ and $1$, respectively. Since the adversary knows the agents’ algorithm, it can pre-compute their movements inside $P_k'$. We now distinguish two cases.

\begin{itemize}
    \item If the agent at $w_{n-1}$ intends to move to $w_n$ via port $1$, the adversary modifies the port assignments of $P_k'$ at the beginning of round $T-1$. Specifically, it connects $w_{n-1}$ to $w_n$ via port $0$ and to the other neighbor via port $1$. Since agents have only $0$-hop visibility and global communication, they cannot detect this change. As a result, the agent takes port $1$ and reaches $w_{n-2}$ instead of $w_n$. Hence, $w_n$ remains unexplored.

    \item If the agent at $w_{n-1}$ either stays put or moves via port $0$, then no agent reaches $w_n$. Hence, $w_n$ again remains unexplored.
\end{itemize}

Thus, regardless of the agents’ actions, node $w_n$ is not explored in round $T-1$. Now suppose that for some round $r \in [T-1, 2T-3]$, node $w_n$ is unexplored. We show that it remains unexplored at round $r+1$. At the beginning of round $r+1$, the adversary keeps $P_1', P_2', \ldots, P_{k-1}'$ unchanged and only modifies $P_k'$. Since $w_n$ is unexplored and there are two agents in $P_k'$ at the end of round $r$, these agents must be at nodes $w_{n-2}$ and $w_{n-1}$. The adversary constructs the new $P_k'$ at round $r+1$ as follows.

\begin{itemize}
    \item If $\beta_r(w_{n-2}) > \beta_r(w_{n-1})$, then it forms $P_k' = w_{n-2} \sim w_{n-1} \sim w_n.$ In this case, $w_{n-1}$ becomes a hole. No agent can reach $w_n$ because it is at least two hops away. Therefore, node $w_n$ remains unexplored at round $r+1$.

    \item If $\beta_r(w_{n-2}) < \beta_r(w_{n-1})$, then it forms $ P_k' = w_{n-1} \sim w_{n-2} \sim w_n.$ In this case, $w_{n-2}$ becomes a hole. Again, $w_n$ remains at distance at least two from any agent. Therefore, node $w_n$ remains unexplored at round $r+1$.

    \item If $\beta_r(w_{n-2}) = \beta_r(w_{n-1}) = 1$, then the adversary forms $ P_k' = w_{n-2} \sim w_{n-1} \sim w_n$. This case is similar to the case discussed at round $T-1$.
    Therefore, the adversary assigns the port numbers using the same pre-computation technique as in round $T-1$. Since the agent at $w_{n-1}$ could not reach $w_n$ at round $T-1$, it also cannot reach $w_n$ at round $r+1$. Therefore, node $w_n$ remains unexplored at round $r+1$.
\end{itemize}

Therefore, $w_n (= v_n)$ is unexplored for all rounds $r \in [T-1, 2T-2]$. By construction, for rounds $r \in [2T-1, 3T-2]$, node $w_n$ becomes isolated again. Using Corollary~\ref{cor:necessary_}, we can always say that in $P_k'$ for round $r\in [iT-1, (i+1)T-2]$, the number of agents are at most two. Since node $w_n(=v_n)$ is unexplored by round $3T-2$, at the beginning of round $3T-1$ in $P_k'$, two agents are at node $w_{n-2}$ or $w_{n-1}$. At round $r=3T-1$, the node $w_n$ remains unexplored because this case is identical to the case for $r\in [T,2T-2]$. Therefore, we can extend the same argument for all $r \geq 3T-1$. Therefore, node $v_n$ remains unexplored for every round $r\geq 0$.

Since capabilities such as global communication, $0$-hop visibility, infinite memory, and complete knowledge of all parameters do not help the agents visit node $v_n$ in any round, our impossibility result holds even under these strong assumptions. This completes the proof.
\end{proof}

\begin{observation}\label{obs:nec}
    By Theorem~\ref{thm:imp_necessary}, solving exploration in Connectivity Time dynamic graphs with $\frac{(n-2)(n-1)}{2}+1$ agents in our model requires 1-hop visibility.
\end{observation}

 \section{Connectivity Time dynamic graph exploration}
 The Connectivity Time dynamic graph has high dynamicity, and the high dynamicity with our port-labelled scheme may prevent even basic coordination if agents are significantly restricted in their ability to perceive their surroundings or communicate with one another. Observation~\ref{obs:nec} implies that, in order to solve the exploration problem with $\frac{(n-2)(n-1)}{2}+1$ agents, the agents must be equipped with $1$-hop visibility. Assume agents have 1-hop visibility. Consider two paths: $P\;=\;w_1\sim\,w_2\sim\,w_3\,\sim w_4\,\sim\,w_5$ and $P'=w_5\sim\,w_2\sim\,w_3\sim\,w_4\sim\,w_1$. The adversary can create either $P$ or $P'$, such that only at $w_3$, the 1-hop view remains unchanged. The movement of agent(s) at $w_3$ remains the same for both $P$ and $P'$, irrespective of the hole position. This can cause exploration to fail. To handle these challenges, we assume that in every round $r$, each agent has 1-hop visibility and global communication. It is important to emphasize that global communication is not fundamental to the definition of the problem but is adopted solely to overcome the adversarial nature of the Connectivity Time model. Moreover, we complement our algorithmic results with impossibility results (refer to Section \ref{sec:imp_Connectivity_Time}) that hold even when agents possess stronger capabilities, including full knowledge of all system parameters, full visibility, and global communication. This highlights the inherent difficulty of exploration in such settings.

\vspace{0.2cm}
\noindent\textbf{High-level idea:} Suppose that at round $r$, agents know the map of $\mathcal{G}_r$. Let $w$ be a multinode and $v$ a hole in $\mathcal{G}_r$, connected via a shortest path $v_1 \sim v_2 \sim \cdots \sim v_p$, with $v_1 = w$, $v_p = v$, and all intermediate nodes $v_2, \ldots, v_{p-1}$ occupied by some agent. Let $a_i$ denote an agent at $v_i$ for $1 \leq i \leq p-1$. In round $r$, each $a_i$ moves to $v_{i+1}$. The multinode $v_1$ remains non-empty, and each $v_i$ ($2 \leq i \leq p-1$) receives an agent from $v_{i-1}$ and sends one to $v_{i+1}$, preserving non-hole status. Finally, the hole $v_p$ is filled. This strategy is known as \emph{pipeline strategy}, which pushes an agent to the nearest hole without creating new holes and has been used in prior works (e.g.,\cite{Ajay_dynamicdisp}).

\begin{figure}
    \centering
    \includegraphics[width=0.5\linewidth]{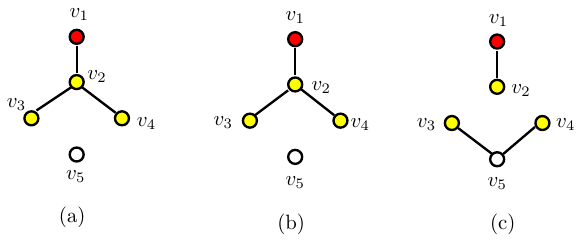}
    \caption{(a) Graph $\mathcal{G}_r$, where $r$(mod 3)=0, (b) Graph $\mathcal{G}_{r}$, where $r$(mod 3)=1, (c) Graph $\mathcal{G}_{r}$, where $r$(mod 3)=2. This figure is an example of the Connectivity Time for $T=3$.}
     \label{fig:exp1}   
\end{figure}

\vspace{0.2cm}
\noindent\textbf{Challenges:} A key challenge arises from the agents’ inability to reconstruct the dynamic graph due to their limited knowledge of the adversary’s behaviour. However, with 1-hop visibility and global communication, agents can share their local views and collaboratively form a partial snapshot of the network. Although this snapshot may not capture the entire graph, it is often sufficient for executing the pipeline procedure. In \cite{Ajay_dynamicdisp}, the authors have used 1-hop visibility and global communication to find a partial map of $\mathcal{G}_r$.

The core difficulty lies in ensuring every node is visited at least once. Even if agents can fill holes, some nodes may remain unvisited. Under the Connectivity Time model, two nodes may never belong to the same connected component in any single round. For example, let $V = \{v_1, v_2, v_3, v_4, v_5\}$ with $T = 3$, where $v_1$ is a multinode, and nodes $v_2, v_3, v_4$ are not holes; node $v_5$ is initially a hole. Define $\mathcal{G}_r = (V, E(r))$ as follows: if $r \bmod 3 = 0$ or $1$, let $E(r) = \{(v_1, v_2), (v_2, v_3), (v_2, v_4)\}$ (see Figure \ref{fig:exp1}(a), (b)), and if $r \bmod 3 = 2$, let $E(r) = \{(v_1, v_2), (v_5, v_3), (v_5, v_4)\}$ (see Figure \ref{fig:exp1}(c)). Although $\mathcal{G}_{r,3} = (V, \bigcup_{i=r}^{r+2} E(i))$ is connected for all $r$, no direct path exists between $v_1$ and $v_5$ in any single round. Thus, $v_1$ cannot pipeline to $v_5$, despite being a multinode. This illustrates how the adversary can isolate nodes round by round, impeding exploration. To address this, we use an \emph{enhanced pipeline}. If at round $r$, agents detect that there exists at least one hole and at least one multinode in their connected component of $\mathcal{G}_r$, then the standard pipeline fills it. If not, then each node in the component has at least one agent. Recall $\alpha_r(u)$ denotes the number of agents at node $u$ at the beginning of round $r$ (refer to Section \ref{sec:model}). If two nodes $u$ and $u_1$ in the same component satisfy $\alpha_r(u_1) \geq \alpha_r(u) + 2$, agents initiate a redistribution along a shortest path $v_1 (=u_1) \sim v_2 \sim \cdots \sim v_p(=u)$ from $u_1$ to $u$, transferring one agent via pipelining to balance the load of agents. Since the graph can be highly sparse and disconnected, the enhanced pipeline gradually builds a configuration of agents on the nodes such that the pipeline becomes feasible along every connected path in each round.

In the following sections, we use the following parameters.

\begin{itemize}
    \item \( \mathcal{A}.\mathit{ID} \): Unique identifier of agent \( \mathcal{A} \).
    
    \item \( \alpha_r(u) \): Number of agents at node $u$ in the beginning of round $r$.
    
    \item \( \pi(v_1, v_2) \): Port label at node \( v_1 \) that leads to its neighbor \( v_2 \).
\end{itemize}

\begin{figure}
    \centering
    \includegraphics[width=0.5\linewidth]{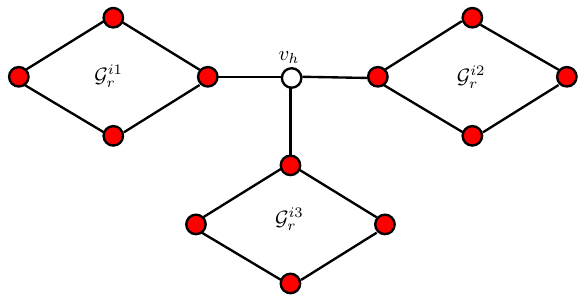}
    \caption{An illustration of the graph $\mathcal{G}_r^i$, where the red-colored nodes contain at least one agent. The node $v_h$ represents a hole.}
    \label{fig:DD1}
\end{figure}

\subsection{Map construction}
In this section, we describe an algorithm that enables the agents to construct a partial map of the graph. Let $\mathcal{G}_r = (V, E(r))$ denote the graph at round $r$. The agents are equipped with global communication and 1-hop visibility. Before presenting the details of the map construction algorithm, we introduce the necessary notation and definitions that will be used throughout its description.

At a round $r$, let $\mathcal{G}_r$ consist of $p$ connected components, denoted by $\mathcal{G}^1_r$, $\mathcal{G}^2_r$, \ldots, $\mathcal{G}^p_r$. Each $\mathcal{G}^i_r=(V_i\, , \, E_i)$ with $V_i\subseteq V$ and $E_i \subseteq E(r)$. Based on the presence of agents at nodes of a connected component, we define the following.

\begin{definition}\label{def:1}
    \textbf{(Connected component of $\mathcal{G}^i_r$ without holes)} Each connected component $\mathcal{G}^i_r$ ($1\leq i \leq p$) can be further divided into subgraphs $\mathcal{G}^{i1}_r$, $\mathcal{G}^{i2}_r$, \ldots $\mathcal{G}^{ik}_r$, where each $\mathcal{G}^{ij}_r \, = \, (V^j_i, E^j_i)$ ($1\leq j \leq k$) satisfies the following:
    
    \begin{itemize}
        \item For every $u \in V^j_i$, at least one agent is present at $u$, $\forall j \in [1,k]$.
        
        \item For all $j \neq l$, $V^j_i \cap V^l_i = \emptyset$ and $E^j_i \cap E^l_i = \emptyset$, $\forall j,l \in [1,k]$.
        
        \item There does not exist an edge $e=(u_1,u) \in E_i$ with $u_1 \in V^j_i$ and $u \in V^l_i$ for $j \neq l$.
    \end{itemize}
\end{definition}

The collection of such subgraphs is denoted by $CC(\mathcal{G}^{i}_r)$. Intuitively, $CC(\mathcal{G}^{i}_r)$ represents the connected components obtained by removing all holes and their associated incident edges from $\mathcal{G}^{i}_r$. As illustrated in Figure~\ref{fig:DD1}, the graph $\mathcal{G}_r^i$ without holes has three connected components, i.e., $|CC(\mathcal{G}_r^i)| = 3$. Now, we proceed with the description of our map finding algorithm, which we call \textsc{MAP()}.

Let agent $\mathcal{A}$ be at a node $v \in \mathcal{G}_r^{i}$. If $\mathcal{A}$ does not have the minimum ID among the agents at node $v$, then it remains idle. Otherwise, the execution of the algorithm \textsc{MAP()} by agent $\mathcal{A}$ in the component $\mathcal{G}_r^{i}$ is divided into two phases at round $r$.

\begin{itemize}
    \item \textbf{Phase 1:} (1-hop view collection) The agent $\mathcal{A}$ performs the following for each port $p \in \{0, 1, \ldots, \deg_r(v) - 1\}$:
    \begin{itemize}
        \item Let $u$ be the neighbor of $v$ reachable via port $p$. Set $ID_v =\mathcal{A}.\mathsf{ID}$. 
        \item If at least one agent is present at $u$: define $C_v^p = (\alpha_r(v),\, ID_v,\, p,\, ID_u)$, where $ID_u$ is the minimum ID among the agents at node $u$.
        \item If no agent is present at $u$ (i.e., it is a hole): define $C_v^p = (\alpha_r(v), ID_v,\, p,\, \bot)$, where $\bot$ denotes a hole via port $p$.
    \end{itemize}
    
    Let $C_v = \{C_v^0,\, C_v^1,\, \ldots,\, C_v^{d}\}$ denote the 1-hop view at node $v$, where $d=\deg_r(v) - 1$. The agent $\mathcal{A}$ broadcasts $C_v$ to all agents in $G_i$ using global communication.

    \item \textbf{Phase 2:} (Graph Reconstruction) Let agent $\mathcal{A}$ receive $k$ distinct $1$-hop views via global communication, denoted by $C_{u_1}, C_{u_2}, \ldots, C_{u_k}$. Agent $\mathcal{A}$ then uses the views $C_{u_i}$ for $i \in [1,k]$, together with its own view $C_v$, to reconstruct a map as follows.

\begin{itemize}
    \item We define \( V' = \{ ID_{u_i} \mid i \in [1,k] \} \cup \{ID_v\} \), where each distinct identifier represents a unique node in the reconstructed graph.

    \item Construct the edge set \( E' \) as follows. For each pair of tuples
\((\alpha_r(w_1),\, ID_{w_1},\, p,\, ID_{w_2})\) and
\((\alpha_r(w_2),\, ID_{w_2},\, q,\, ID_{w_1})\),
where \( ID_{w_1} \neq \bot \), \( ID_{w_2} \neq \bot \), and
\( w_1, w_2 \in \bigcup_{i=1}^k \{u_i\} \cup \{v\} \),
add an undirected edge \((ID_{w_1}, ID_{w_2})\) with port labels
\( \pi(ID_{w_1}, ID_{w_2}) = p \) and
\( \pi(ID_{w_2}, ID_{w_1}) = q \).

For each tuple \((\alpha_r(w_1),\, ID_{w_1},\, p,\, \bot)\), mark port \( p \) at node \( ID_{w_1} \) as leading to a hole.\footnote{This step does not introduce any new holes or edges incident to holes during Phase~2; it only records the ports at node \( w_1 \) that lead to a hole.}

\end{itemize}
\end{itemize}

\begin{theorem}\label{thm:map}
Let agent $\mathcal{A}$ be at a node $v$ in the connected component $\mathcal{G}_r^i$. Then, by executing the subroutine \textsc{MAP()}, agent $\mathcal{A}$ reconstructs $CC(\mathcal{G}_r^i)$ together with the port information indicating which ports of nodes in $CC(\mathcal{G}_r^i)$ lead to holes in $\mathcal{G}_r^i$.
\end{theorem}

\begin{proof}
Let $(w_1, w_2)$ be an edge in $CC(\mathcal{G}_r^i)$ for some $i$, and let $\pi(w_1, w_2) = p$ and $\pi(w_2, w_1) = q$. Since $(w_1, w_2)$ belongs to $CC(\mathcal{G}_r^i)$, both $w_1$ and $w_2$ contain at least one agent. Let $ID_{w_1}$ and $ID_{w_2}$ denote the minimum IDs among the agents located at nodes $w_1$ and $w_2$, respectively. Hence, agent $\mathcal{A}$ receives the $1$-hop views $C_{w_1}$ and $C_{w_2}$, and consequently $ID_{w_1}, ID_{w_2} \in V'$. Because $\pi(w_1, w_2) = p$ and $\pi(w_2, w_1) = q$, we have $C_{w_1}^p \in C_{w_1}$ and $C_{w_2}^q \in C_{w_2}$. By the construction of \textsc{MAP()}, these tuples are
\[
C_{w_1}^p = (\alpha_r(w_1), ID_{w_1}, p, ID_{w_2}) \quad \text{and} \quad
C_{w_2}^q = (\alpha_r(w_2), ID_{w_2}, q, ID_{w_1}).
\]
Therefore, agent $\mathcal{A}$ adds an undirected edge $e = (ID_{w_1}, ID_{w_2})$ to $E'$ with port labels $\pi(ID_{w_1}, ID_{w_2}) = p$ and $\pi(ID_{w_2}, ID_{w_1}) = q$. Hence, $e \in E'$.

Now consider a node $w_1 \in CC(\mathcal{G}_r^i)$ such that one of its ports, say $p$, leads to a hole. Since $w_1 \in CC(\mathcal{G}_r^i)$, it contains at least one agent. Let $ID_{w_1}$ be the minimum ID among the agents at $w_1$. Then agent $\mathcal{A}$ receives $C_{w_1}$. Because port $p$ of $w_1$ leads to a hole, we have
\[
C_{w_1}^p = (\alpha_r(w_1), ID_{w_1}, p, \bot).
\]
By the rules of \textsc{MAP()}, agent $\mathcal{A}$ correctly records that port $p$ of node $ID_{w_1}$ leads to a hole.

Therefore, agent $\mathcal{A}$ reconstructs all edges of $CC(\mathcal{G}_r^i)$ and correctly identifies all ports of its nodes that lead to holes in $\mathcal{G}_r^i$. This completes the proof.
\end{proof}

From Theorem \ref{thm:map} and Definition \ref{def:1}, we derive three key observations.

\begin{observation}\label{obs:info_map}
By executing \textsc{MAP()}, the agents in $\mathcal{G}_r^i$ reconstruct $CC(\mathcal{G}_r^i)$, including the information about the number of agents at each node $v$ and the ports of $v$ (if any) that lead to holes.
\end{observation}

\begin{observation}\label{obs:nohole_map}
If there is no hole in $\mathcal{G}_r^i$, then $CC(\mathcal{G}_r^i)$ coincides with $\mathcal{G}_r^i$, as per Definition~\ref{def:1}.
\end{observation}

\subsection{Perpetual exploration algorithm}
In this section, we present the algorithm \textsc{EXP\_ALGO()}, which solves the perpetual exploration problem when agents have 1-hop visibility and global communication. The following is a detailed description of the algorithm for agent $\mathcal{A}$ at node $v$ at round $r$.

If agent $\mathcal{A}$ is not the minimum ID agent at node $v$, then it stays at node $v$ at round $r$. Otherwise, agent $\mathcal{A}$ executes the algorithm \textsc{MAP()} and constructs a graph $G'$. The agent then considers the connected component of $G'$ that contains its current position $v$, denoted by $G''$. Depending on the structure of $G'$, one of the following four cases may arise, and in each case, agent $\mathcal{A}$ proceeds as described below.

\begin{itemize}
     \item \textbf{Case 1 (there is no multinode and no information of a port towards a hole in $G'$):} Agent $\mathcal{A}$ stays at node $v$.

     \item \textbf{Case 2 (there is no multinode in $G'$, but there is information of a port leading to a hole):}  Let $u_1, u_2, \ldots, u_k$ be the nodes in $G'$ such that for each $j \in [1,k]$, there is at least one agent at node $u_j$, and one of the ports of $u_j$ leads to a hole. Let $b_j$ denote the agent with the minimum ID at node $u_j$, for each $j \in [1,k]$. If $\mathcal{A}.ID=\text{min}\{b_j.ID: j\in[1,k] \}$, then it moves to the node which leads to the hole via the minimum port. Otherwise, it stays at its position. 

     \item \textbf{Case 3 (both a multinode and information of a port towards a hole are present in $G'$):} Let $u_1, u_2, \ldots, u_k$ be the nodes in $G'$ such that for each $j \in [1,k]$, node $u_j$ is a multinode. Let $b_j$ denote the agent with the minimum ID at node $u_j$, for each $j \in [1,k]$. Without loss of generality, let $b_1.ID=\text{min}\{b_j.ID: j\in[1,k] \}$. 

     \begin{itemize}
         \item If $u_1\in G''$: Let $\overline{u}_1, \overline{u}_2, \ldots, \overline{u}_{k'}$ be the nodes in $G''$ such that for each $j \in [1,k']$, there is at least one agent at node $\overline{u}_j$, and one of the ports of $\overline{u}_j$ leads to a hole. Let $\overline{a}_j$ denote the agent with the minimum ID at node $\overline{u}_j$, for each $j \in [1,k']$. Without loss of generality, let $\overline{a}_1.ID=\text{min}\{\overline{a}_j.ID: j\in[1,k'] \}$.
        
        Since agent $\mathcal{A}$ is aware of $G''$, it consider a shortest path $P$ between $u_1$ and $\overline{u}_1$. If there are multiple shortest paths between $u_1$ and $\overline{u}_1$, then it selects the one that is lexicographically shortest among all other shortest paths. Let $P=w_1(=u_1)\sim w_2\sim \ldots \sim w_y(=\overline{u}_1)$ be the lexicographically shortest path in $G''$. If $v=w_j$ for $1\leq j<y$, it moves to node $w_{j+1}$. Else if agent $\mathcal{A}$ is at node $w_y$, it moves to the node via the minimum available port, which leads to a hole. Otherwise, it stays at node $v$.
         \item If $u_1\notin G''$: It stays at its position.
     \end{itemize} 
        
        \item  \textbf{Case 4 (there is a multinode in $G'$ but there is no information of a port towards a hole in $G'$):} As per Observation \ref{obs:nohole_map}, graphs $G'$ and $G''$ are the same map. Assume there are $k$ nodes in $G''$. Let $V(G'')$ be the set of nodes in $G''$. Define: $i' = \max\{ \alpha_r(x) : x \in V(G'') \}$, and $j' = \min\{ \alpha_r(x) : x \in V(G'') \}$. Let \( v_1, v_2, \ldots, v_p \) be the nodes in \( G'' \) such that \( \alpha_r(v_k) = i' \) for each \( k \in [1, p] \), and let $\overline{a}_k$ denote the minimum ID agent at node $v_k$. Similarly, let \( w_1, w_2, \ldots, w_q \) be the nodes in $G''$ such that \( \alpha_r(w_{k'}) = j' \) for each \( k' \in [1, q] \), and let $\overline{b}_{k'}$ denote the minimum ID agent at node $w_{k'}$. Without loss of generality, let $\overline{a}_1$ be the agent among $\{\overline{a}_k : k \in [1, p]\}$ with the minimum ID, and $\overline{b}_1$ be the agent among $\{\overline{b}_{k'} : k' \in [1, q]\}$ with the minimum ID. If $\alpha_r(v_1)< \alpha_r(w_1)+2$, agent $\mathcal{A}$ stays at node $v$. Otherwise (i.e.,$\alpha_r(v_1)\geq \alpha_r(w_1)+2$), agent $\mathcal{A}$ finds a shortest path $P$ between $v_1$ and $w_1$. If there are multiple shortest paths between $v_1$ and $w_1$, then it selects the one that is lexicographically shortest. Let $P=z_1(=v_1)\sim z_2 \ldots \sim z_y(=w_1)$. If $z_j=v$ for some $1\leq j<y$, it moves to node $z_{j+1}$. Otherwise, it stays at $v$.
\end{itemize}

In the next section, we show the correctness of our algorithm.

\subsection{Correctness and analysis of the algorithm}\label{sec:correctness}
Before proving correctness, we introduce the following notation. Let $n$ be the number of nodes, and define $
l = \frac{(n-2)(n-1)}{2} + 1.
$ For each $i \in [0, l]$, let $
S_i := \{ v \in V : \alpha_r(v) = i \}.
$ Let $L$ denote the largest integer such that $S_L \neq \emptyset$ at round 0.

\begin{lemma}\label{lm:cor_hole}
    Let $\mathcal{G}_r$ be the configuration at round $r$. If there exists at least one hole and at least one multinode in the same connected component of $\mathcal{G}_r$, say $\mathcal{G}_r^i$, then the total number of holes in $\mathcal{G}_r$ decreases by at least one at the end of round $r$. 
\end{lemma}
\begin{proof}
At round $r$, let the cardinality of set $CC(\mathcal{G}_r^i)$ is $p$ where $p\geq 1$. As per Observation \ref{obs:info_map}, the agents in $\mathcal{G}_r^i$ reconstruct $CC(\mathcal{G}_r^i)$ using \textsc{MAP()}, including the information about the number of agents at each node $v$ and the ports of $v$ (if any) that lead to holes. Therefore, the map $G'$ constructed by any agent in $\mathcal{G}_r^i$ is  $CC(\mathcal{G}_r^i)$. According to Definition~\ref{def:1},
$CC(\mathcal{G}_r^i)=\big\{\mathcal{G}_r^{i1}, \mathcal{G}_r^{i2}, \ldots, \mathcal{G}_r^{ip}\big\}$ for some $p$. Let $u_1, u_2, \ldots, u_k$ be the nodes in $G'$ such that for each $j \in [1,k]$, node $u_j$ is a multinode. Let $b_j$ denote the agent with the minimum ID at node $u_j$, for each $j \in [1,k]$. Without loss of generality, let $b_1.ID=\text{min}\{b_j.ID: j\in[1,k] \}$. As per \textsc{EXP\_ALGO()}, agents in $\mathcal{G}_r^{ij}$, for $j\in [2,p]$ stays at its position. Only agents of $\mathcal{G}_r^{i1}$ execute the algorithm further in this case. For agents in $\mathcal{G}_r^{i1}$, the map $G''$ is nothing but the graph of $\mathcal{G}_r^{i1}$ including which port leads to a hole. Since there is a hole in $\mathcal{G}_r^i$, there is a port from a node of $\mathcal{G}_r^{i1}$ which leads to a hole. Let there be $k''$ such nodes, denoted $\overline{u}_1, \overline{u}_2, \ldots, \overline{u}_{k''}$, 
with corresponding minimum-ID agents $\overline{a}_1, \overline{a}_2, \ldots, \overline{a}_{k''}$. Let $\overline{a}_1.ID = \min\{\overline{a}_j.ID : j \in [1, k'']\}$. 

Since all agents in $\mathcal{G}_r^{i1}$ construct the same map $G''$, they identify the same ordered pair of nodes: $u_1$ (a multinode) and $\overline{u}_1$ (a node adjacent to a hole). Let $P = (v_1 = u_1 \sim v_2 \sim \ldots \sim v_y = \overline{u}_1)$ denote the lexicographically shortest path from $u_1$ to $\overline{u}_1$ in $G''$. This path is unique due to deterministic selection based on agent IDs and their common knowledge of $G''$. Each agent on this path determines its position and acts accordingly. If an agent $\mathcal{A}$ is at node $v_i$ for $i < y$ and is the minimum ID agent at $v_i$, it moves to $v_{i+1}$. The agent at $v_y = \overline{u}_1$ selects the minimum available port leading to a hole, say $v_h$, and moves through it. Thus, an agent occupies the hole node $v_h$ by the end of round $r$.

Therefore, the total number of holes in $\mathcal{G}_r$ decreases by at least one at the end of round $r$. This completes the proof.
\end{proof}

\begin{lemma}\label{lm:cor_nohole}
Let $\mathcal{G}_r$ be the configuration at round $r$, and suppose that there exists a connected component $\mathcal{G}_r^i$ of $\mathcal{G}_r$ such that it contains at least one multinode but no hole. Define $
i' = \max\{ \alpha_r(x) : x \in V_i \} \text{ and } j' = \min\{ \alpha_r(x) : x \in V_i) \}$, where $V_i$ is the set of nodes in $\mathcal{G}_r^i$. If $i' \geq j' + 2$, then at the end of round $r$, one of the nodes $v \in V_i$ with $\alpha_r(v) = i'$ reduces its $\alpha_r$ by 1, and one of the nodes $w \in V_i$ with $\alpha_r(w) = j'$ increases its $\alpha_r$ by 1.
\end{lemma}
\begin{proof}
    At round $r$, let $\mathcal{G}_r$ consist of $p$ connected components, denoted $\mathcal{G}_r^1, \mathcal{G}_r^2, \ldots, \mathcal{G}_r^p$. Without loss of generality, let $\mathcal{G}_r^i$ be the connected component that contains at least one multinode but no hole. Since $\mathcal{G}_r^i$ contains a multinode but no hole, by Theorem~\ref{thm:map} and Observation~\ref{obs:nohole_map}, the map $G'$ constructed by each agent using \textsc{MAP()} is exactly $\mathcal{G}_r^i$. Due to Observation \ref{obs:nohole_map}, the map of $G''$ is nothing but the map of $G'$. Let \( v_1, v_2, \ldots, v_p \) be the nodes in \( \mathcal{G}_r^1 \) such that \( \alpha_r(v_k) = i' \) for each \( k \in [1, p] \), and let $a_k$ denote the minimum ID agent at node $v_k$. Similarly, let \( w_1, w_2, \ldots, w_q \) be the nodes such that \( \alpha_r(w_{k'}) = j' \) for each \( k' \in [1, q] \), and let $b_{k'}$ denote the minimum ID agent at node $w_{k'}$. Without loss of generality, let $a_1$ be the agent among $\{a_k : k \in [1, p]\}$ with the minimum ID, and $b_1$ be the agent among $\{b_{k'} : k' \in [1, q]\}$ with the minimum ID. Since all agents share the same reconstructed map $\mathcal{G}_r^i$, they all identify the same pair of nodes $v_1$ and $w_1$ as the nodes with maximum and minimum $\alpha_r$ values, respectively. Let $P = (v_1 = u_1 \sim u_2 \sim \cdots \sim u_y = w_1)$ be the lexicographically shortest path from $v_1$ to $w_1$ in $\mathcal{G}_r^1$, which is uniquely determined by the map and agent ID choices. Each agent on this path identifies its position and moves accordingly: if an agent is at node $u_i$ for $i < y$, it moves to $u_{i+1}$. As a result, the value of $\alpha_r(v_1)$ decreases by 1, and the value of $\alpha_r(w_1)$ increases by 1. This completes the proof.
\end{proof}

\begin{remark}\label{rk:1}
    Based on Lemma \ref{lm:cor_nohole}, we observe that one agent reaches a node $w_1 \in S_{j'}$ at round $r$. As a result, $\alpha_r(w_1)$ becomes $j'+1$ at the beginning of round $r+1$, implying that $w_1 \in S_{j'+1}$. 
    Whenever at round $r$, a node becomes a part of $S_p$, i.e., at round $r-1$, it was not a part of $S_p$, we call this event \emph{join}. Whenever at round $r$, a node does not remain a part of $S_p$, i.e., at round $r-1$, it was a part of $S_p$ but it is not part of $S_p$ at round $r$, we call this event \emph{leave}.
\end{remark}

 We have an observation based on $\mathsf{EXP\_ALGO()}$, which is as follows.

\begin{observation}\label{obs:hole}
A node $v$ with $\alpha_r(v) \geq 1$ at round $r$ can become a hole by the end of round $r$ only if $\alpha_r(v) = 1$ and $v$ belongs to a connected component $\mathcal{G}_r^i$ of $\mathcal{G}_r$ that contains a hole but no multinode.
\end{observation}

We now show that perpetual exploration is achieved using $l$ agents.
\begin{lemma}\label{lm:|S_0|=1}
    If $|S_0|= 1$ at some round $r$, then node $v\in S_0$ is visited by some agent within the next $T$ rounds.
\end{lemma}
\begin{proof}
    To maintain the Connectivity Time property, node $v$ must be connected to some node $w$ at round $t$, where $t\in [r, \, r+T]$. Let $\mathcal{G}_t^1$ be the connected component of $\mathcal{G}_t$ at round $t$ such that the node $v$ is part of the graph $\mathcal{G}_t^1$. If $\mathcal{G}_t^1$ has at least one multinode, then one agent moves to node $v$ as agents in $\mathcal{G}_t^1$ execute $\mathsf{EXP\_ALGO()}$, and due to Lemma \ref{lm:cor_hole}, the number of holes decreases by 1.

    Otherwise, all nodes in $\mathcal{G}_t^1$ except node $v$ contain exactly one agent. At round $t$, let $w_1$, $w_2$, \ldots, $w_p$ be neighbours of node $v$ in $\mathcal{G}_t^1$, and agent $a_i$ be at node $w_i$. As per $\mathsf{EXP\_ALGO()}$, the minimum ID agent ID among $a_i$s moves to node $v$ as there is only one hole in $\mathcal{G}_t$. This completes the proof.
\end{proof}

\begin{lemma}\label{lm:Condition}
    If $|S_0|\geq 2$, then $\exists \;i, \,j \,(\geq i) \in [0, l]$ such that $S_i \neq \emptyset$, $S_j \neq \emptyset$, $j \geq i+2$ and $S_k = \emptyset$ for all $i < k < j$.
\end{lemma}
\begin{proof}
Suppose the lemma does not hold. It implies $|S_i| \geq 1$ for every $i\in [1,L]$, where $L$ is the largest index satisfying $S_L \neq \emptyset$. The value $L$ is $\leq n-2$. Assume $L>n-2$. Without loss of generality, let $L=n-1$. Since $|S_i|\geq 1$ for every $i\in [1, L]$, $\sum_{i=1}^L|S_i|\geq n-1$. Therefore, the total number of nodes is $|S_0|+\sum_{i=1}^L|S_i|\geq 2+n-1=n+1$ as $|S_0|\geq 2$ and $\sum_{i=1}^L|S_i|\geq n-1$. This leads to the contradiction as $n$ many nodes are present. Therefore, $L\leq n-2 $.

Since $|S_i|\geq 1$ for every $i\in[1,L]$, $L\leq \sum_{i=1}^L|S_i|\leq n-2$. Therefore, $L\leq n-2$. Define $X = \sum_{i=1}^{L} i \cdot |S_i|$. The value $X$ denotes the total number of agents. The maximum value of $X$ occurs when $|S_i| = 1$ for $1 \leq i \leq L-1$, and $|S_L| = n - |S_0| - (L-1)\leq n-L-1$ as $|S_0|\geq 2$. Thus, $X\leq 1 + 2 + \dots + (L-1) + L \cdot (n - L - 1)=L\cdot (L-1)/2 \,+\, L\cdot(n-L-1)$. After simplifying, we get that $X\leq L\cdot(2n-L-3)/2\leq (n-2)(n-1)/2$. This leads to the contradiction as the number of agents in the system is $\frac{(n-2)(n-1)}{2}+1$. Thus, our initial assumption must be incorrect. This completes the proof.
\end{proof}

\begin{lemma}\label{lm:correct_S_0>1}
    If initially $|S_0|\geq 2$, then within $O(n^4\cdot T)$ round $|S_0|\leq 1$. 
\end{lemma}
\begin{proof}
    At round $0$, there are two possible cases: Case 1: $|S_1| = 0$, or Case 2: $|S_1| \geq 1$.

    \vspace{0.2cm}
    \noindent \textbf{Case 1:} To maintain the Connectivity Time, some node from $S_0$ must be in the same connected component as a node from $S_i$, $i \geq 2$, within the first $T$ rounds. By Lemma \ref{lm:cor_hole}, this causes $|S_0|$ to decrease by at least one in the next $T$ rounds. If $|S_1| = 0$ throughout $[0, n\cdot T]$, then $S_0$ eventually becomes empty. Otherwise, if $|S_1| \geq 1$ at some round $r \in [0, n\cdot T]$, we are in Case 2.
    
     \vspace{0.2cm}
    \noindent \textbf{Case 2:} Suppose that at round \( r \geq 0 \), $|S_1|\geq 1$. To show that \( |S_0| \leq 1 \), we proceed by contrapositive argument. Assume that \( |S_0| \geq 2 \) throughout the interval $[r, r + (n^4+1) \cdot T]$. Then, by Lemma~\ref{lm:Condition}, there exist indices \( i \) and \( j \) with \( j \geq i + 2 \) such that \( S_i \neq \emptyset \), \( S_j \neq \emptyset \), and \( S_k = \emptyset \) for all \( k \in [i+1, j-1] \).

        According to Remark \ref{rk:1}, nodes may join or leave the set $S_p$ at each time step. A key question is: how many times can nodes join the set $S_p$? By Lemma \ref{lm:cor_nohole}, a node can join $S_p$ at round $t \geq r$ only if at least one node from \( S_q \) (with \( q \geq p+1 \)) and one node from \( S_{p-1} \) belong to the same connected component \( \mathcal{G}_t^{i_1} \) of \( \mathcal{G}_t \), and no node from \( S_{p'} \) for \( p' < p-1 \) is in \( \mathcal{G}_t^{i_1} \). We have \( l\) many agents. Then in the worst case, nodes can join the set \( S_p \) at most \( l \) times, as each time one agent can move from \( S_q \) to \( S_{p-1} \). If at most $l$ times nodes joins set $S_p$, the size of \( S_q \) decreases due to Lemma \ref{lm:cor_nohole}, eventually making \( S_q = \emptyset \) for every \( q > p \).

        As there are $L$ such sets with $L \leq l$, the total number of join events between rounds $r$ and $r + (n^4 + 1) \cdot T$ can not be more than $l^2 \leq n^4$. Now divide the interval $[r, r + (n^4 + 1) \cdot T]$ into \( n^4 + 1 \) consecutive sub-intervals of length \( T \): each sub-interval is of the form $[r + kT, r + (k+1)T - 1]$ for $k = 0, 1, \dots, n^4+1$. If at least one node leaves some set $S_j$ during each sub-interval, then there are $n^4 + 1$ such leave events in total. Since each leave corresponds to a join, this implies there are also $n^4 + 1$ join events. But this contradicts our earlier conclusion that there can be at most \( n^4 \) join events. By the pigeonhole principle, this contradiction implies that our assumption is false. Therefore, within $O(n^4 \cdot T)$ rounds, we must have \( |S_0| \leq 1 \).

 Since Case 2 is reached within $O(n \cdot T)$ rounds from any initial configuration, $|S_0| \leq 1$ holds within $O(n^4 \cdot T)$ rounds overall. This completes the proof.
\end{proof}

\begin{theorem}\label{th:main}
$\mathsf{EXP\_ALGO()}$ solves perpetual exploration in Connectivity Time dynamic graphs using \(\frac{(n-2)(n-1)}{2} + 1 \) synchronous agents equipped with 1-hop visibility, global communication, and \( O(\log n) \) bits of memory. The agents have no knowledge of \( n \), \( T \), \( l \).
\end{theorem}
\begin{proof}
If \( |S_0| \geq 2 \) initially, then by Lemma~\ref{lm:correct_S_0>1}, \( |S_0| \leq 1 \) within \( O(n^4 \cdot T) \) rounds. Suppose that at some round \( r \geq 0 \), \( |S_0| \leq 1 \). If \( S_0 = \emptyset \), then all nodes are occupied, and by Observation~\ref{obs:hole}, no node becomes a hole in future rounds. Thus, perpetual exploration is achieved.

If \( S_0 = \{v\} \), then by Lemma~\ref{lm:|S_0|=1}, node \( v \) is visited within the next \( T \) rounds. Thus, all nodes are visited by some agent within \( O(n^4 \cdot T) \) rounds, and the invariant \( |S_0| \leq 1 \) is preserved thereafter due to Observation \ref{obs:hole}. This process repeats indefinitely, ensuring perpetual exploration.

Each agent requires \( O(\log n) \) bits to distinguish itself. Since the computation is round-local (i.e., agents do not retain information from previous rounds), \( O(\log n) \) bits of memory are sufficient. This completes the proof.
\end{proof}

\begin{remark}
    Lemma \ref{lm:correct_S_0>1} shows that at least $n-1$ nodes are occupied in the first $O(n^4 \cdot T )$ rounds. Theorem \ref{th:main} ensures that from this point onward, at most one hole exists and, by Lemma \ref{lm:|S_0|=1}, it is revisited within each$\,\,T$ rounds. Therefore, this also implies that each node of the network is visited by some agent in the first $O(n^4 \cdot T)$ rounds.
\end{remark}

\section{Conclusion and future work}\label{sec:conclusion}
In this work, we studied the exploration problem in two dynamic graph models, namely $1$-Interval Connectivity and Connectivity Time. For the $1$-Interval Connectivity model, we established an impossibility result that closes an existing gap in the assumptions on agents’ visibility and communication capabilities. For the Connectivity Time model, we proved that exploration is impossible with $\frac{(n-2)(n-1)}{2}$ agents starting from an arbitrary initial configuration, even when agents are equipped with strong capabilities such as infinite memory, full visibility, global communication, and complete knowledge of all parameters. We further showed that $\frac{(n-2)(n-1)}{2}+1$ agents are sufficient to solve exploration provided that the agents have $1$-hop visibility, and we presented an explicit algorithm that achieves this using $1$-hop visibility, global communication, and only $O(\log n)$ memory.

Despite these results, our algorithm does not guarantee termination even when agents have $1$-hop visibility and global communication. This raises several important open questions. In particular, it remains unclear whether termination can be guaranteed under these assumptions, and whether global communication is truly necessary to solve exploration with $\frac{(n-2)(n-1)}{2}+1$ agents. Investigating these questions, as well as understanding finer trade offs between visibility, communication, and memory, constitutes a promising direction for future research.

\section{Acknowledgement}
Ashish Saxena would like to acknowledge the financial support from IIT Ropar. Kaushik Mondal would like to acknowledge the ISIRD grant provided by IIT Ropar. This work was partially supported by the FIST program of the Department of Science and Technology, Government of India, Reference No. SR/FST/MS-I/2018/22(C). We acknowledge the reviewers of DISC 2025 for their careful reading and insightful comments, which contributed to improving the presentation and quality of this paper.

\vspace{0.4cm}
\noindent\textbf{Declaration of generative AI and AI-assisted technologies in the writing process} 

During the preparation of this work, we used \emph{Grammarly, ChatGPT} tools in order to improve language quality. After using this tool, we reviewed and edited the content as needed and take full responsibility for the content of the publication.

\bibliography{bib}

\end{document}